%% file: main-sigmod.tex
\documentclass[sigconf]{acmart}

\usepackage{amsmath}
\usepackage{algorithm}
\usepackage[noEnd=true, indLines=true]{algpseudocodex}
\usepackage{xspace}
\usepackage{graphicx}
\usepackage{tabularx}
\usepackage{subcaption}
\usepackage{multicol}
\usepackage{multirow}
\usepackage[normalem]{ulem}
\usepackage{xcolor}

\usepackage{enumitem}

\AtBeginDocument{%
  }

\setcopyright{acmlicensed}
\acmJournal{PACMMOD}
\acmYear{2025} \acmVolume{3} \acmNumber{6 (SIGMOD)}
\acmArticle{315} \acmMonth{12} \acmPrice{}
\acmDOI{10.1145/3769780}

\theoremstyle{theorem}
\newtheorem{definition}{Definition}
\newtheorem{theorem}{Theorem}

\newcommand{\stitle}[1]{\vspace*{0.4em}\noindent{\bf #1.\/}}

\newcommand{\squishlist}{
	\begin{list}{$\bullet$}
		{ \setlength{\itemsep}{1pt}
			\setlength{\parsep}{1pt}
			\setlength{\topsep}{2.5pt}
			\setlength{\partopsep}{0.5pt}
			\setlength{\leftmargin}{1em}
			\setlength{\labelwidth}{1em}
			\setlength{\labelsep}{0.6em}
		}
	}
	\newcommand{\squishend}{
	\end{list}
}

\newcommand{\name}{{DiskJoin}\xspace}

\definecolor{darkyellow}{RGB}{204, 153, 0}
\definecolor{darkgreen}{RGB}{0, 128, 0}
\newcommand{\rone}[1]{#1}
\newcommand{\rthree}[1]{#1}
\newcommand{\rfour}[1]{#1}

\begin{document}

\title{DiskJoin: Large-scale Vector Similarity Join with SSD}

\author{Yanqi Chen}
\affiliation{%
  \institution{University of Massachusetts Amherst}
  \country{}
}
\email{yanqichen@cs.umass.edu}

\author{Xiao Yan}
\affiliation{%
  \institution{Centre for Perceptual and Interactive Intelligence}
  \country{}
}
\email{yanxiaosunny@gmail.com}

\author{Alexandra Meliou}
\affiliation{%
  \institution{University of Massachusetts Amherst}
  \country{}
}
\email{ameli@cs.umass.edu}

\author{Eric Lo}
\affiliation{%
  \institution{Chinese University of Hong Kong}
  \country{}
}
\email{ericlo@cse.cuhk.edu.hk}

\begin{abstract}
  Similarity join---a widely used operation in data science---finds all pairs of items that have distance smaller than a threshold. 
   Prior work has explored distributed computation methods to scale similarity join to large data volumes but these methods require a cluster deployment, and efficiency suffers from expensive inter-machine communication.
   On the other hand, disk-based solutions are more cost-effective by using a single machine and storing the large dataset on high-performance external storage, such as NVMe SSDs, but in these methods the disk I/O time is a serious bottleneck. 
   In this paper, we propose \name, the first disk-based similarity join algorithm that can process billion-scale vector datasets efficiently on a single machine. 
   \name improves disk I/O by tailoring the data access patterns to avoid repetitive accesses and read amplification. It also uses main memory as a dynamic cache and carefully manages cache eviction to improve cache hit rate and reduce disk retrieval time. For further acceleration, we adopt a probabilistic pruning technique that can effectively prune a large number of vector pairs from computation. Our evaluation on real-world, large-scale datasets shows that \name significantly outperforms alternatives, achieving speedups from 50$\times$ to 1000$\times$. 
\end{abstract}




\maketitle

\input{sections/1-introduction}
\input{sections/2-related-work}
\input{sections/3-solution-overview}
\input{sections/4-efficiency-optimizations}

\input{sections/5-evaluation}
\input{sections/6-conclusion}

\bibliographystyle{ACM-Reference-Format}
\bibliography{references}

\end{document}

%% file: sections/1-introduction.tex
\section{Introduction}\label{sec:intro}

In machine learning workflows, it is common practice to use models to map objects with complex semantics (e.g., text, images, videos, proteins) to vector embeddings and manage these objects via embeddings~\cite{kiela2014learning, naseem2020transformer, vasile2016meta}. For instance, convolution neural networks (CNNs) map images to embeddings~\cite{kiela2014learning}, transformer-based large language models (LLMs) map text segments to embeddings~\cite{naseem2020transformer}, and recommendation models map users and products to embeddings~\cite{vasile2016meta}. On these vector embeddings, similarity---e.g., measured via Euclidean distance or cosine similarity---is a crucial notion. For example, close similarity of two vector embeddings is used to indicate that two images are similar, that two text segments have similar meanings, or that a user may like a product.

\begin{figure}[!t]
	\begin{tabular}{lccc}
		\toprule
		\textbf{Metric}    &    & \textbf{DiskANN~\cite{diskann}} & \textbf{\name [this paper]} \\ 
		\midrule
		Total time (sec)   & & $130,107$  & $\textbf{1,185}$        \\
		Disk time (sec)    & & $93,417$ & \textbf{198}                         \\
		Disk traffic (GB)  & & $115,854$   & $\textbf{265}$            \\
		\bottomrule
	\end{tabular}
    \vspace{-2mm}
    	\caption{
        Profiling results for a baseline solution---using the state-of-the-art SSD-based vector index DiskANN~\cite{diskann} to perform vector similarity join---and our \name on the BigANN100M dataset at 90\% recall. Both methods use a memory size that is 10\% of the dataset size.}
        	\label{tab:introstat}
\end{figure}

\stitle{Vector similarity join} 
\rone{Similarity self-join (SSJ), which seeks all pairs of similar data points, is a well-established problem that has been extensively studied in the literature. SSJ for vector data lies at the core of many applications like near-duplicate detection~\cite{semdedup,video-dedup,document-dedup}, outlier detection~\cite{anomaly}, and more.}
We give three practical examples and then provide a formal problem definition.
\begin{description}[topsep=0pt, leftmargin=6pt, labelsep=0pt]
    \item \emph{Video deduplication}~\cite{dedup}: A video website maintains a large number of videos, but many are similar to each other~\cite{similar-video, video-redundancy}. For example, the same clip may be uploaded by different users or encoded in different formats; highly-overlapping segments may be cropped from one video. Identifying and removing these duplicates can save  resources~\cite{near-duplicate} and avoid repetitive content recommendations. 
    Similar videos (as well as other formats like images or text) can be identified by performing SSJ over their vector embeddings.

    \rone{\item \emph{Training data deduplication}~\cite{semdedup}: Dataset deduplication is a crucial step in machine learning pipelines, especially when training large language models (LLMs), where the training datasets often contain billions of examples. In particular, semantic deduplication---which identifies pairs of data that are semantically similar but not exactly identical---has been shown to significantly reduce dataset size, while maintaining, or even improving, model performance. A practical approach to enabling semantic deduplication at scale is to apply similarity self-join over vector embeddings of the data.}
    
    \item \emph{Outlier detection}~\cite{chandola2009anomaly}: An object is considered an outlier if it significantly deviates from other objects. With the embedding vectors, the deviation can be quantified using the number of neighbors that are within a distance threshold from a vector~\cite{outlier}. If a vector has only a few similar neighbors, we can conclude that its corresponding object is an outlier, which requires special attention (e.g., a fraudulent transaction or abnormal training sample).  

\end{description}


\begin{definition}[Similarity self-join (SSJ) for vector dataset]\label{def:problem}
	
Given a vector dataset $\mathcal{X}=\{x_1, x_2,\cdots, x_N\}$, where each $x_n \in \mathcal{X}$ ($1 \!\le\! n \!\le\! N$) is a $d$-dimensional vector, and a threshold $\epsilon>0$, find all vector pairs $(x_n, x_{n'})$ with $\Vert x_n- x_{n'} \Vert\le \epsilon$ and $1\le n\neq n' \le N$.
	
\end{definition}

In this paper (including Definition~\ref{def:problem}), we use the L-2 norm (Euclidean distance) as the similarity measure, but other similarity measures are also possible. 
The similarity-based self-join (SSJ) problem may be extended to cross-join if we consider similar vector pairs $(x_n, y_m)$ from two vector datasets $\mathcal{X}$ and $\mathcal{Y}$. 
We assume that the vector dimension $d$ is high (e.g., $>100$), which is the common case for embedding models. 
Note that solving the SSJ problem exactly (i.e., finding all qualifying vector pairs with $\Vert x_n- x_{n'} \Vert\le \epsilon$) is prohibitively expensive. This is because traditional distance-based pruning techniques (e.g., using triangle inequality) are not effective in high dimensions~\cite{curse-of-dimensionality}, and thus the problem almost degrades to a linear scan over the dataset $\mathcal{X}$ to find all $\epsilon$-neighbors of each vector $x_n$. As such, we solve SSJ \textit{approximately} to trade accuracy for efficiency.  We can gain efficiency by reducing the number of candidate vector pairs that are checked for $\epsilon$-neighbors; this reduction may miss some vector pairs that should have been returned, thus hurting recall.  We use the standard definition for recall: If $\mathcal{R}$ is the set of all vector pairs whose distance is less than $\epsilon$, then the recall of an approximate result set $\mathcal{R}'$ of vector pairs is: $r=\frac{|\mathcal{R} \cap \mathcal{R}'|}{|\mathcal{R}|}$.



\rone{
Existing literature (see Section~\ref{sec:related-work} for a detailed discussion), treats SSJ as an \emph{end-to-end task over a static dataset} executed exclusively by a dedicated server or system, with the goal of minimizing the overall execution time. For example, RSHJ~\cite{lshsj} conducts approximate similarity join using LSH-based probing, where the entire join pipeline---from hashing to candidate generation to filtering---is optimized for overall performance. Similarly, distributed similarity join solutions like MAPSS~\cite{mapss}, ClusterJoin~\cite{clusterjoin}, and C2Net~\cite{c2net} partition and process data in a way that tightly integrates filtering and verification stages, focusing on reducing total job runtime.

Our work is aligned with this line of research: we also treat SSJ as a unified, end-to-end task over a static dataset.
}
Thus, our objective is to gain efficiency (measured by end-to-end execution time), while maintaining high recall.
We do not need to separately consider precision, as incorrect candidate pairs with distances larger than  $\epsilon$ will be simply removed by checking their distances. Therefore, an approximate solution to SSJ would always achieve perfect precision. 

\stitle{Scaling to large data sizes} 
In many applications, SSJ often involves large datasets. For example, a public multi-modal dataset can provide billions of images and their text descriptions~\cite{laion}, a video website such as YouTube can also host billions of videos~\cite{youtube}, and an e-commerce platform can have hundreds of millions of users~\cite{ecommerce}. Moreover, each individual vector may also be large due to the high dimensionality $d$. As a result, vector datasets can take up TBs of storage and do not fit in the memory of a server. A way to scale up SSJ is to use distributed methods that involve multiple machines. However, distributed solutions are expensive to deploy and usually suffer from costly inter-machine communication~\cite{mapss,clusterjoin}. 

\stitle{SSD-based alternative and challenges} SSDs can offer a practical and cost-effective alternative to distributed SSJ: they are cheap (1TB currently costs $\sim$\$60), capacious (10+ TBs are common), and reasonably fast (reaching 1.4M random IOPS and 7 GB/s bandwidth).  
An alternative storage option is persistent memory (PM), which is byte-addressable, persistent, and offers SSD-like capacity at a lower cost than DRAM. However, PM is significantly more expensive (\$12/GB) while delivering comparable read bandwidth (9 GB/s vs 7.4 GB/s) but lower write bandwidth (2.8 GB/s vs 6.9 GB/s). Although PM offers much lower latency (300ns vs 40µs), this advantage has minimal impact on end-to-end applications like similarity join.
Thus, we study the case where the vector dataset is stored on SSD, while the memory is only a small fraction (e.g., $<$ 20\%) of the data.


\looseness-1
A baseline disk-based SSJ approach is to use disk-oriented vector indexes (e.g., DiskANN~\cite{diskann}), which target top-$k$ nearest neighbor queries: we first build an index on the data and then use each vector as a query to search the index. We gradually increase $k$ until the distances of the identified neighbors exceed $\epsilon$. Figure~\ref{tab:introstat} shows that this approach is slow and disk access dominates the overall time, due to the large volume of disk traffic. Our analysis suggests that the poor performance boils down to two reasons: \emph{read amplification} and \emph{repetitive access}. In particular, disk-oriented indexes access an individual vector each time but a vector is usually smaller than the 4KB disk page size (e.g., a 128-dimension float vector takes up 512B), which is the minimum granularity of disk read.\footnote{While some more recent embeddings have high dimensionality (e.g., 1536 for OpenAI's text embedding), dimensionality reduction techniques like the Matryoshka Representation Learning (MRL)~\cite{mrl} and vector quantization methods~\cite{lvq,gao2024practical} can be used to compress the vector size to well below a page size without loss of accuracy.}
Thus, disk bandwidth is wasted on reading irrelevant data. Critically, this approach treats the vectors as independent queries and misses the opportunities to share and reuse the data fetched from disk. For example, vectors $x_a$ and $x_{b}$ may both need to check if $x_{c}$ is their $\epsilon$-neighbor. If $x_{b}$ is processed right after $x_a$, $x_{c}$ will  reside in memory and can be reused, leading to smaller disk traffic and shorter runtime.

\stitle{Our solution: \name} 
Motivated by this profiling analysis, we design a novel disk-based SSJ method, \name, that effectively reduces disk access time and, in turn, significantly improves the efficiency of SSJ processing. \name  achieves this with two key designs: \emph{access batching} and \emph{task orchestration}. 

\emph{Access batching} addresses read amplification by reading sets of vectors (buckets), instead of one vector at a time.
At a high level, \name organizes vectors in the dataset into buckets; each bucket is represented by a center vector and holds the vectors that are similar to the center vector. Using the bucket centers, \name then constructs a \emph{bucket graph}, where each bucket is a node, and an edge connects two buckets if vectors in one bucket may have $\epsilon$-neighbors in the other bucket.  Our method evaluates SSJs by processing one bucket at a time, retrieving all neighbors of that bucket in the bucket graph.  By reading data at bucket granularity, instead of individual vectors, \name largely eliminates read amplification, as each bucket contains multiple vectors and typically has size larger than the page size.  It is also conducive to sharing disk access, as vectors in the same bucket are similar, and thus are likely to have $\epsilon$-neighbors in common.  We make further system contributions through careful implementation to ensure that memory consumption stays below budget when assigning the vectors to buckets, and that construction of the bucket graph is fast and incurs limited overhead.


\emph{Task orchestration} further reduces disk access through more effective sharing of disk access and better-informed cache eviction policy. 
It achieves this by choosing a processing order for buckets, to leverage that buckets processed consecutively can share the buckets loaded to memory (i.e., temporal locality).
Identifying the optimal bucket ordering is NP-hard (Section~\ref{sec:method:problem}).  However, we leverage a popular algorithm for classical graph ordering (which assigns IDs to graph nodes, such that nodes with adjacent IDs share common neighbors), as an approximation heuristic, and process buckets in the order of their new IDs.
We note that for a given bucket processing order, the bucket graph provides complete knowledge of the required bucket accesses. We choose Belady's algorithm~\cite{Belady} to manage the memory cache eviction as it is proven to maximize the cache hit rate given perfect knowledge of future data access.   

\stitle{Similarity join vs vector similarity search}
We note that similarity self-join has different requirements from vector similarity search (VSS)~\cite{nsg, hnsw, ivf}. VSS is an \emph{online task} with stringent latency requirements for each query (e.g., 10ms). 
VSS cannot collect many queries within the latency budget period, and thus the opportunities for the joint optimizations of multiple queries are limited. In contrast, SSJ is an \emph{offline batch processing task} that focuses on end-to-end time. As such, \name batches the processing of vectors that check the same $\epsilon$-neighbors and obtains the data access pattern beforehand for orchestration. These designs make \name fundamentally different from disk-based indices for VSS~\cite{diskann,starling}.   

\rone{
Some recent research explores \emph{filtered} vector similarity search by integrating attribute filtering into the index. For instance, Filtered-DiskANN~\cite{filtered-diskann} builds a graph whose edges respect both geometric proximity and shared attribute labels. SeRF~\cite{serf} overlays range-specific subgraphs to support filtering over value ranges. NHQ~\cite{nhq} constructs a composite graph index to enable joint pruning on vector and attribute similarity. We are, however, not aware of any research on filtered vector similarity join. While our work focuses on the standard setting of vector SJ---without attribute filtering---we briefly discuss how our approach can be extended to support attribute filtering in Section~\ref{sec:overview}.
}

\stitle{Summary of evaluation and results}
We evaluate \name on three real-world datasets that contain up to a billion vectors and limit the memory size to a small portion of the dataset size (e.g., 10\%). The results show that compared with searching a disk-oriented vector index, \name is 2--3 orders of magnitude faster. As suggested by the results in Figure~\ref{tab:introstat}, this is because \name significantly reduces disk traffic and hence disk access time. Moreover, Figure~\ref{tab:introstat} shows that disk access no longer dominates the running time for \name. Our experiment suggests that the designs of \name are effective at improving cache hit rate and reducing disk access. Specifically,  Belady's algorithm and bucket ordering improve the cache hit rate by about 20\% and 50\%, respectively.   

\stitle{Contributions and outline}
We make the following contributions.
\begin{itemize}[topsep=0pt, leftmargin=6pt]
    \item We organize a review of the related work on similarity joins, highlighting the existing bottlenecks of huge disk traffic and long disk access time. [Section~\ref{sec:related-work}]
    \item We provide the intuition and overview for our novel \name method for SSJ.  \name reduces disk access drastically, through clever implementation of bucket-wise processing of SSJs.  Through the key contributions of access batching and task orchestration, \name allows for more sharing of disk data access and richer information for more effective cache management. [Section~\ref{sec:overview}]
    \item We model the problem of task orchestration as an edge-covering problem of the bucket dependency graph and show theoretical results on its hardness. We proceed to provide practical heuristics that leverage prior work on cache management and graph ordering, in a careful, memory-conscious implementation that performs task orchestration efficiently. [Section~\ref{sec:method}]
    \item We detail the optimizations that support the effectiveness of our bucketization model, allowing for clustering the vectors with limited memory and  pruning unnecessary vector similarity computations. [Section~\ref{sec:opt}]
    \item We experimentally evaluate our methods over three real-world datasets of up to 1.4B vectors, and show that \name offers clear, consistent, and up to 3-orders-of-magnitude improvement over the prior state of the art. [Section~\ref{sec:eval}]
\end{itemize}






%% file: sections/2-related-work.tex
\section{Related Work}\label{sec:related-work}

Similarity join (both self-join and cross-join) is a classical problem that has been extensively studied for data types including strings~\cite{xiao2008ed, qin2011efficient, SSJoin, adaptfilter, MGjoin}, sets~\cite{SSJoin, GPjoin, adaptfilter, arasu2006efficient}, and vectors~\cite{c2lsh, c2net, ego, ego-star, gorder, eD-tree, R-tree, M-tree}. The general solution is a two-step \textit{filter-verify} procedure, where the \textit{filter} step generates a set of candidate pairs, and the \textit{verify} step checks the candidate pairs via similarity computation. While the \textit{verify} step is straightforward, the \textit{filter} step is crucial for accuracy and efficiency by excluding dissimilar pairs and keeping similar pairs. Many filtering techniques have been proposed, including those targeting generic metric spaces~\cite{R-tree, M-tree, eD-tree, mapss, clusterjoin, chen2016metric} as well as techniques tailored to specific data types~\cite{xiao2008ed, arasu2006efficient, c2lsh}.      


\stitle{Similarity join for strings and sets} 
Prefix filtering is a dominant technique for set and string similarity joins. Consider sorted sets $r$ and $s$, and denote their length-($|r|-t+1$) prefixes as $\mathsf{P}(r, |r|-t+1)$ and $\mathsf{P}(s, |r|-t+1)$; if $r$ and $s$ have more than $t$ common elements, it holds that $\mathsf{P}(r, |r|-t+1)\cap \mathsf{P}(s, |r|-t+1)\neq \emptyset$. Using this property, SSJoin~\cite{SSJoin} generates candidate pairs by considering objects whose prefixes have common elements, and can handle both sets and strings. AdaptJoin~\cite{adaptfilter} improves prefix filtering by selecting the first $|r|-t+l$ elements as the prefix and filtering out object pairs whose prefixes share fewer than $l$ common elements; $l$ is selected to balance the costs of filtering and verification. GPJoin~\cite{GPjoin} groups identical prefixes for sets and executes filtering in batches. 
MGJoin~\cite{MGjoin} reduces the number of candidates by using multiple prefixes with different lengths for each string during filtering.



\stitle{Similarity join for low-dimensional vectors} Space-partitioning indices can prune dissimilar vectors in low  dimensions (e.g., $d<10$). For example, tree-based variants, such as R-tree~\cite{R-tree}, $M$-tree~\cite{M-tree} and $eD$-tree~\cite{eD-tree}, partition the space hierarchically. Search for $\epsilon$-neighbors recurses from the root and uses the triangle inequality to prune tree branches that are unlikely to contain qualified neighbors. Some algorithms use the $\epsilon$-grid (e.g., EGO~\cite{ego}, EGO-star~\cite{ego-star}, SuperEGO~\cite{super-ego}), which splits the space into $\epsilon$-sized grids and sorts the vectors by the lexicographical order of their grid coordinates. Then, the vectors are grouped into parts, and two parts are checked if their $\epsilon$-margin bounding boxes intersect. Space-partitioning indices do not work for high-dimensional vectors (e.g., $d>100$) produced by machine learning models, as the pruning power of the triangle inequality and bounding boxes diminishes in high dimensions~\cite{curse-of-dimensionality}. 


\begin{figure*}[!t]
	\centering
	\includegraphics[width=\textwidth]{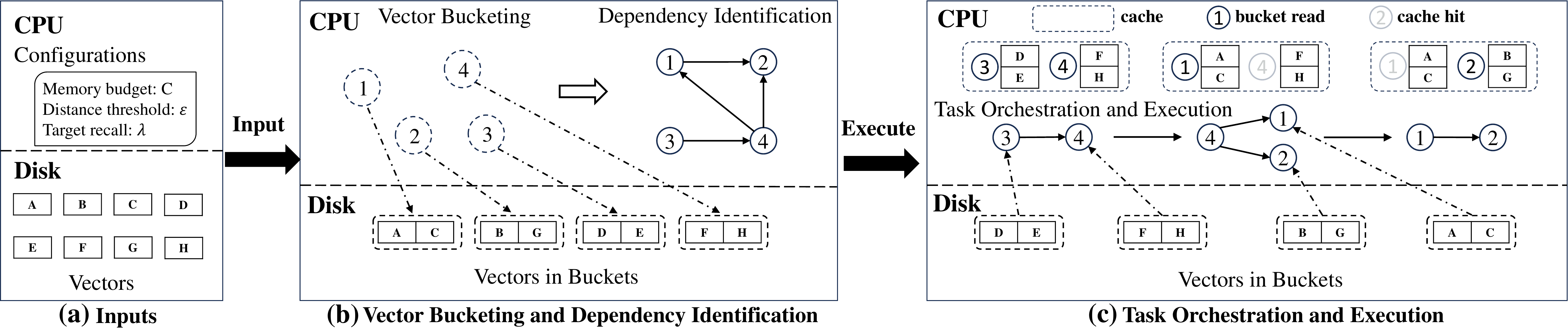}
    \vspace{-6mm}
	\caption{Workflow of \name. Numbers (e.g., 1 and 2) indicate  buckets, and letters (e.g., A and B) indicate  vectors. (a) The inputs are the vector dataset and task configurations; (b) the vectors are grouped into buckets, and a bucket graph is constructed, where an edges means that the vectors in one bucket needs to check another bucket for neighbors; (c) task orchestration decides a good processing order for the edges in the bucket graph to reduce cache miss, where the cache can hold 2 buckets. 
    }
	\label{fig:sys-overview}
\end{figure*}

\looseness-1
\stitle{Similarity join for high-dimensional vectors} RSHJ~\cite{lshsj} conducts similarity cross-join for two vector datasets $\mathcal{X}$ and $\mathcal{Y}$ and uses locality-sensitive hashing (LSH)~\cite{qalsh}, a popular technique for vector similarity search in high dimensions. It assumes that $\mathcal{X}$ and $\mathcal{Y}$ reside in memory and produces an approximation (i.e., identifies most, instead of all similar pairs) like we do. LSH uses tailored hash functions to map vectors to buckets ensuring that similar vectors are more likely to be mapped to the same bucket. RSHJ  builds hash tables for both datasets and  generates candidate pairs by querying the hash table of $\mathcal{Y}$ using vectors in $\mathcal{X}$. To reduce lookup cost, it only uses a representative subset of $\mathcal{X}$ for lookups.  However, state-of-the-art vector indices, such as proximity graphs~\cite{diskann, hnsw, hcnng} and  inverted file (IVF)~\cite{ivf}, perform much better than LSH for similarity search.

\stitle{Distributed similarity join} Some prior work scales up similarity join with distributed frameworks like MapReduce~\cite{mapreduce}.  MAPSS~\cite{mapss} first selects a set of centroids and then partitions the dataset according to the centroids. The mappers use the triangle inequality to filter the candidate objects for each partition, and each reducer verifies the candidate pairs for a partition. ClusterJoin~\cite{clusterjoin} improves MAPSS by pruning the candidate pairs with a bisector-based filter in general metric spaces, as well as a set of filters specific to different distance functions.  It achieves load balancing by re-partitioning the datasets over the machines. Chen et al.~\cite{chen2016metric} further enhance candidate filtering using pivot filtering and plane sweeping techniques and propose two partitioning methods based on space-filling curve and KD-tree to avoid re-partitioning. More recently, C2Net~\cite{c2net} applies an LSH variant, C2LSH~\cite{c2lsh}, to conduct similarity-based vector cross-join by hashing both datasets to buckets. The mappers enumerate the vector pairs mapped to the same bucket, and the reducers aggregate the pairs to calculate their collision counts. A vector pair is regarded as a candidate if their collision count exceeds a threshold. 
These distributed solutions generally suffer from a high inter-machine communication cost because similar object pairs need to be sent to the same machine for verification.

\looseness-1
To the best of our knowledge, \name is the first to conduct similarity join for high-dimensional vectors on disk. Compared with existing vector similarity join work (e.g., EGO, RSHJ, C2Net) that use tree, grid, or LSH, \name leverages state-of-the-art vector indices (i.e., proximity graph and IVF), which are more efficient for high-dimensional vectors. Compared with distributed solutions, \name is easier to deploy (as it does not require a cluster) and does not suffer from high inter-machine communication cost. \name also follows the general \emph{filter-verify} procedure for similarity join but the main challenge is to reduce the disk access cost, which has not been explored by existing work. 

\rthree{
\stitle{Disk-based Similarity Search}
Prior work has explored disk‑based vector similarity search systems to support billion‑scale vector retrieval with low memory footprints and high throughput. GRIP~\cite{grip} pioneered memory-SSD hybrid similarity search by introducing a multi-store IVF-based framework, where the in-memory index quickly routes queries to nearby clusters and completes validation by accessing the SSD store. SPANN~\cite{spann} adopts similar design but optimizes the IVF index construction to better suit the disk scenario. DiskANN~\cite{diskann} proposes a graph-based approach, using compressed vectors for in-memory graph traversal and refining results with full-precision vectors stored on SSDs. Other systems, such as SmartANNS~\cite{smartanns} and ES4D~\cite{es4d}, also explore the opportunity for near-data processing, which offloads the query processing to SSDs equipped with computational capabilities.

\stitle{Batch Processing}
\name leverages batch processing, which is a widely-used concept in database research.
SharedDB \cite{shareddb} first introduced the design of batching queries in RDBMSs to share data access and computation. This concept was later extended to vector similarity search in~\cite{sanca2024efficient}. Query batching is especially common in GPU-based similarity search systems---such as GGNN~\cite{ggnn}, SONG~\cite{song}, and BANG~\cite{bang}---to fully exploit the massive parallelism of GPUs. Other systems, including FreshDiskANN~\cite{freshdiskann} and IP-DiskANN~\cite{ipdiskann}, also employ batch processing for index updates to amortize operational overhead.
}

%% file: sections/3-solution-overview.tex
\section{\name  Overview}\label{sec:overview}


In this section, we provide an overview of \name's workflow, depicted in  \autoref{fig:sys-overview}. 
\name takes as input the vector dataset $\mathcal{X}$, the memory budget $C$, the distance threshold $\epsilon$ for similar vector pairs, and a target recall $\lambda$. 
Under the assumption that $\mathcal{X}$ resides on local disk, \name will identify and write a set of similar vector pairs $\mathcal{R}'$ to disk, such that the expected recall of $\mathcal{R}'$ is $\lambda$ and memory usage is bounded by $C$. 
\rthree{The core of \name is the bucket-wise processing of SSJ. It first groups $\mathcal{X}$ into buckets, such that each bucket contains similar vectors. Then, it identifies candidate bucket pairs that may contain neighboring vectors. Finally, for each candidate bucket pair, it performs pairwise vector comparisons across the buckets to produce the final results.} 
For this approach to be effective, the algorithm needs to be smart about the bucketization, the identification of likely dependencies across buckets, and the processing order of buckets.
This provides three benefits: (1)~it avoids read amplification, as the vectors are read from disk at bucket granularity; (2)~data access is shared among the vectors in the same bucket because they are similar and hence are likely to have high overlap in their $\epsilon$-neighbors;  (3)~understanding the dependencies across buckets (i.e., which buckets are likely to contain $\epsilon$-neighbors) allows for advanced memory cache management. We provide here an overview of these key components in \name, and describe them in detail in subsequent sections.

\stitle{Vector bucketization} 
\name partitions $\mathcal{X}$ into buckets, with each bucket $b$ represented by a center vector $c_b$ and containing similar vectors. Traditional clustering methods, such as k-means, are not suitable for this task, due to high memory requirements---often, they need the entire dataset to fit in memory, but, in practical settings, the memory size $C$ is much smaller than the dataset size.
\name implements an efficient bucketization procedure (detailed in Section~\ref{sec:opt:bucketing}).  At a high level, it starts with a random selection of vectors 
to serve as the bucket centers, and then assigns each vector to the nearest center. 
Each bucket also records a radius $r_b$, which is the maximum distance between its vectors and the center. 

\name also adjusts the disk layout of the dataset by storing the vectors of each bucket consecutively (\autoref{fig:sys-overview}b). This allows for fetching each bucket via sequential disk reads.  A naive implementation of reorganizing the disk layout that uses random reads to retrieve the vectors for each bucket and then writes them back to disk would cause read amplification.  We instead implement a more efficient, sequential-disk-access method that does not violate the memory budget (Section~\ref{sec:opt:bucketing}).
   
\setlength{\belowdisplayskip}{2pt} \setlength{\belowdisplayshortskip}{1pt}
\setlength{\abovedisplayskip}{2pt} \setlength{\abovedisplayshortskip}{1pt}

\stitle{Dependency identification} 
\name constructs a \emph{bucket graph} that indicates, for each bucket $b$, which other buckets contain vectors that need to be checked for potential $\epsilon$-neighbors.
In particular, after vector bucketization, we have the center and radius of the buckets in memory (i.e., $c_b$ and $r_b$ for bucket $b$).
Using the triangle inequality, bucket $b_j$ may contain $\epsilon$-neighbors of bucket $b_i$ if: 
\begin{equation}\label{equ:trangle}
	\Vert c_{b_i}-c_{b_j}\Vert-r_{b_i}-r_{b_j}\le \epsilon.
\end{equation}

Since the triangle inequality has poor pruning power in high-dimension spaces, each bucket will have many candidate neighboring buckets. Inspired by~\cite{pruning}, we adopt the probabilistic bucket pruning rule (Section~\ref{sec:opt:pruning}), which meets the target recall $\lambda$ while pruning many more candidate buckets. We denote the bucket graph as $G(V,E)$, where each bucket $b\in V$ is a node, and a directed edge $e_{ij}\in E$ if 
bucket $b_i$ and $b_j$ are a candidate bucket pair and $i<j$. We do not record edges with $i>j$ because Euclidean distance is symmetric, i.e., by computing the distances of the vectors in $b_i$ to the vectors in $b_j$, we also obtain the distances from $b_j$ to $b_i$.

\stitle{Task orchestration} 
The edges of the bucket graph identify all pairs of buckets that need to be processed together to identify $\epsilon$-neighbors. A \emph{task} refers to the processing of an edge 
(i.e., bucket pair).  SSJ completes when all edges are processed.  The order of processing the edges is critical for cache hit rate, which in turn determines IO cost and impacts efficiency, as a bucket needs to be loaded from disk upon cache miss. For example, if edge $e_{jh}$ is processed after $e_{ij}$, it will not incur a cache miss for bucket $b_j$, as $b_j$ will already reside in memory.  We define the problem of identifying the bucket load sequence to process all edges with minimum cache misses and show that it is NP-hard (Section~\ref{sec:method:problem}).  We then propose a heuristic approximation that decomposes it into two sub-problems: optimal cache management and task ordering.  For a given task order, the bucket graph provides complete knowledge of future data access; thus, we use Belady's algorithm~\cite{Belady} for cache management (Section~\ref{sec:method:belady}), which has been proven to yield the minimum total cache misses when data access is known.  To solve task ordering, we leverage \emph{graph reordering} methods. The intuition is that buckets that have a large overlap in their neighboring buckets should be processed consecutively for temporal cache locality (Section~\ref{sec:method:reorder}).

\stitle{Task execution} 
Task orchestration generates the processing order of the edges and the buckets to load to and evict from the cache memory at every step. During execution, 
our method evaluates the distances of vector pairs from two buckets each time via in-memory computation to check the $\epsilon$-neighbors for the vectors in each bucket, and follows the task orchestration to load and evict buckets. The resulting $\epsilon$-similar vector pairs are written to disk.

\stitle{Extending to cross-join} \name can be extended to process the cross-join for two vector datasets with the following changes. (1)~Vector bucketing is conducted separately for the two datasets, generating their own buckets. (2)~The bucket graph becomes a bipartite graph, with the buckets of one dataset requiring the buckets of the other dataset. 
(3)~During task orchestration, we use graph reordering to decide the processing order of the buckets from the larger dataset, the memory cache to hold buckets from the smaller dataset, and Belady's algorithm to manage cache eviction. The other way, i.e., reordering the smaller dataset and caching the larger dataset, is less efficient. This is because the reordered dataset only needs to be streamed to the memory once while the cached dataset will experience bucket loading and eviction. Our experiments show that the disk traffic can be several times the dataset size.           

\rone{
\stitle{Extending to attribute filtering} 
\name can be extended to support attribute filtering with a simple modification: before performing pairwise vector comparisons between candidate bucket pairs, apply the attribute filter to each vector to generate a bitmap, and skip similarity computations for vectors that do not pass the filter. If the attribute filter satisfies the range-preserving property---that is, the filtered results form a contiguous block in the lexicographical order of attribute values---further optimization is possible. Specifically, during vector bucketization, vectors within each bucket can be sorted by the attributes' lexicographical order. As a result, during task execution, only the vectors that satisfy the filter need to be loaded from disk, using a single sequential read.
}


%% file: sections/4-efficiency-optimizations.tex
\section{Task Orchestration}\label{sec:method}

In this section, we describe how \name determines the processing order of bucket pairs and how it handles cache management during execution.  At this stage, we assume a known bucket graph; we describe the bucketization and graph construction in Section~\ref{sec:opt}.  We start by defining the problem of covering all graph edges with minimum cache misses, and show that it is NP-hard.  We then propose an approximation heuristic that decomposes it into the sub-problems of cache management (Section~\ref{sec:method:belady}) and task ordering (Section~\ref{sec:method:reorder}).


\subsection{Edge Covering with Cache Constraints}\label{sec:method:problem}

\looseness-1
We assume a bucket graph $G=(V,E)$, where each node $v\in V$ is a bucket, and an edge $(u,v)\in E$ indicates that buckets $u$ and $v$ may contain vectors that are $\epsilon$-close. To compute the distance of vectors in the bucket pair $(u,v)$, both $u$ and $v$ have to be in memory. We assume that all buckets are of the same size, and, with slight abuse of notation, we use $C$ to denote the number of buckets that fit in memory. To perform SSJ, we need to process all edges of the bucket graph, and the goal is to minimize the disk access cost, which corresponds to minimizing the number of bucket load operations. We define the \emph{minimum edge cover with cache} (MECC) problem as follows.


\begin{definition}[MECC: Minimum Edge Cover with Cache]\label{def:edge-cover}
Assume a given undirected graph $G=(V,E)$ and an initially-empty cache with capacity $C$. The cache supports two operations: $\mathsf{load}(v)$, which loads node $v$ to the cache, and $\mathsf{evict}(v)$, which evicts node $v$ from the cache following a given policy $P$ when the cache is full and needs to load a node. An edge of $(u,v)\in G$ is said to be covered if both $u$ and $v$ are present in the cache at the same time. The Minimum Edge Cover with Cache problem seeks to find the sequence of load operations with the minimum length that covers all edges of $G$.
\end{definition}

We note that edge direction in bucket graphs simply implies that each pair of buckets needs to be processed once, but directionality is not relevant to the covering problem. Thus, we consider $G$ to be undirected in this discussion. Without loss of generality, we assume that $G$ is connected---disconnected components can simply be handled independently of each other.  The length of the node load sequence corresponds to the number of buckets loaded from disk to memory, and thus MECC minimizes the disk access cost.  The minimum possible length of the load sequence is $|V|$, as each node needs to be loaded at least once. 
\autoref{example:edge-cover} shows an example of MECC, where the graph has 4 nodes, the cache capacity is 2, and the length of the minimum load sequence is 7. The MECC in \autoref{example:edge-cover} is easy to solve because two subsequently-checked edges share at most 1 bucket. 
However, MECC becomes difficult in general cases. 

\begin{theorem}\label{theorem:Np-hard}
The MECC problem is NP-hard.
\end{theorem}
\begin{proof}
We show that the decision version of MECC can be reduced from independent set (IS), which, given a graph $G=(V,E)$ and an integer $T$, asks whether there exists a set $V'\subseteq V$ such that $|V'|= T$ and $\not\exists u,v\in V'$ such that $(u,v)\in E$.
Given an instance of IS with connected graph $G(V,E)$ and integer $T$, we construct an instance of MECC with the same graph $G(V,E)$ and a cache of size $C=|V|-T+1$ with last-in-first-out (LIFO) as the eviction policy $P$.   

We first show that if $G$ has an independent set of size $T$, the solution of MECC is a load sequence of length $|V|$.  Let  $V_T\subseteq V$ be an independent set of size $T$ in $G$. We create a load sequence of length $|V|$ that loads the nodes $V\backslash V_T$ first into the cache, followed by the nodes in $V_T$. When executing this load sequence, nodes $V\backslash V_T$ will take up $|V|-T$ slots in the cache, leaving one free slot.  Each subsequent node in $V_T$, is loaded into that slot, and, after being processed, gets evicted (based on LIFO) to load the next node.  Since the nodes in $V_T$ form an independent set, there are no edges between them. All edges in $E$ are either among nodes in $V\backslash V_T$, which are together in the cache, and thus are covered, or between a node in $u\in V_T$ and nodes in $V\backslash V_T$, which are covered when $u$ is loaded.  Thus, this sequence covered all edges in $G$.

\begin{figure}[!t]
	\centering
        \includegraphics[width=0.47\textwidth]{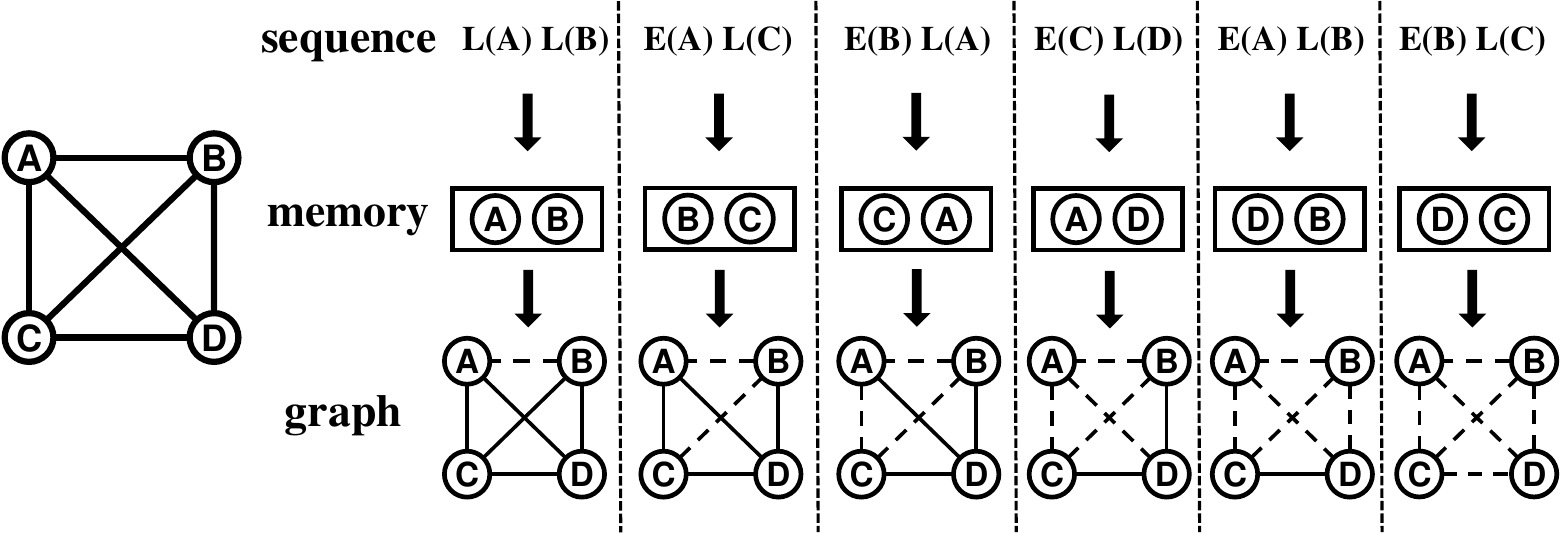}
    \vspace{-3mm}
	\caption{An example of the minimum edge covering with cache (MECC) problem, the graph is in the left plot, the cache size is 2, and $L(\cdot)$ and $E(\cdot)$ mean load and evict a node, respectively. The dotted lines show the edges covered in each step.}
	\label{example:edge-cover}
	\Description{}
\end{figure}

\begin{figure}[!t]
    \centering
    \includegraphics[width=\columnwidth]{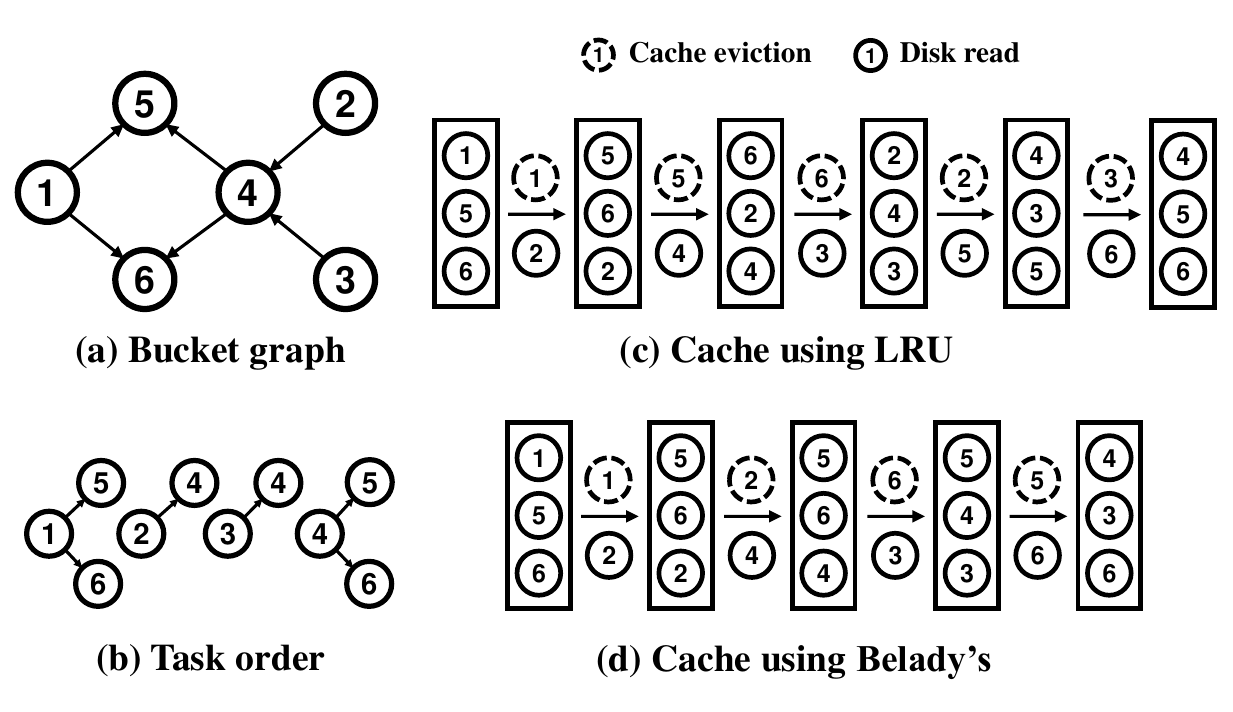}
    \vspace{-6mm}
    \caption{An example for Belady's algorithm. The edge processing order is in (b), the cache size is 3, LRU loads 8 buckets while Belady's algorithm loads 7 buckets.}
    \label{fig:belady}
\end{figure}

Conversely, we also show that if the MECC instance can be solved with exactly $|V|$ node loads (so every node is loaded once) and a cache size of $|V|-T+1$ using LIFO, the set of evicted nodes form an independent set of size $T$. This is because any two evicted nodes are never present in the cache at the same time, and if any edges existed between them they would not have been covered. 

Since IS is NP-hard, and IS reduces to MECC in polynomial time, it follows that  MECC is also NP-hard.
\end{proof}

We note that this hardness result is specific to self-joins.  Cross-joins are likely simpler, since their bucket graph is bipartite.  Self-joins are generally harder, and our algorithms can also be used in the evaluation of cross-joins, as discussed in Section~\ref{sec:overview}.  However, it is possible that cross-joins offer additional opportunities for improvements; we leave this direction to future work.

As MECC is NP-hard, we solve it approximately by decomposing it into two sub-problems.
\squishlist 
\item \emph{Given a processing order of the edges, we examine the cache eviction policy that minimizes the number of bucket loads.} We show that this can be solved optimally using Belady's algorithm for cache management in Section~\ref{sec:method:belady}. 

\item \emph{We examine how to derive a good processing order for the edges to achieve a small number of bucket loads.} We solve this heuristically, using a lightweight graph reordering algorithm in Section~\ref{sec:method:reorder}.

\squishend

\subsection{Cache Management via Belady's Algorithm}\label{sec:method:belady}


Assume that the processing order for the edges in $G(V,E)$ has been decided as $\{e_{uv},\cdots,e_{u'v'}\}$, we can deduce the access sequence of the buckets as $S=\{u, v, \cdots, u', v'\}$ by extracting the buckets connected by each edge. $S$ has a length of $2|E|$ and specifies when each bucket will be used for computation. The cache can keep $C$ buckets in memory. Given the bucket access sequence, our goal is to minimize the number of bucket loads with cache management. 

\looseness-1
The cache load decision is straightforward, i.e., when checking edge $e_{uv}$ and a required bucket (e.g., $u$) is not in the cache, the bucket should be fetched from disk. However, when $u$ needs to be loaded and the cache is full, we need to choose which bucket to evict from the cache. There are popular strategies such as first-in-first-out (FIFO), least-frequently-used (LFU), and least-recently-used (LRU), with the latter often being the choice method in many practical settings. However, these policies make eviction decisions based on \emph{prior} data access.  What is special in our setting is that the bucket access sequence $S$ specifies all data access, including the future. Thus, we use Belady's algorithm for cache management, which can leverage this information to minimize the number of cache misses~\cite{Belady}. 

Belady's algorithm looks into future bucket access and evicts the bucket in the cache that will be accessed again at the latest point in the future. This minimizes the number of cache misses because, compared with the bucket chosen for eviction, the other buckets will be accessed sooner; thus, evicting any of these buckets would yield at least the same number of cache misses as evicting the chosen bucket. Figure~\ref{fig:belady} compares LRU with Belady's algorithm with an example. LRU~(c) evicts bucket $b_5$, which needs to be loaded again later to process another edge. In contrast, Belady's algorithm~(d) evicts bucket $b_2$ because it will never be used again, and this keeps $b_5$ in the cache and saves one future bucket load.  

\begin{algorithm}[!t]
	\caption{Cache Management via Belady's Algorithm} 
	\label{alg:belady}
    {\small
	\begin{algorithmic}[1]
		\Require Bucket read sequence $S$, bucket number $M$, cache size $C$
		\State Bucket queue $\mathsf{Q} \leftarrow  \textsf{Max\_Heap}(C)$  
		\State Bucket position $P \leftarrow$ Array($M$, List())
		\State Bucket access count $c \leftarrow$ Array($M$, 0)
		\For{$i\in \{1\cdots |S|\}$} \label{ln:initP} \Comment{Store access time for each bucket}
		\State $P[S[i]].\mathsf{append}(i)$ \label{ln:initP2}
      \EndFor
		\For{$i\in \{1\cdots |S|\}$} \label{ln:main1} \Comment{Enumerate bucket access sequence}
		\State $b \leftarrow S[i]$, \ $c[b] \leftarrow c[b] + 1$
		\If{$\mathsf{Cache.find}(b) = \mathsf{true}$} \Comment{Cache hit}
		\State $\mathsf{Q.update}(b, P[b][c[b]])$ \label{ln:cacheHit}
		\Else \Comment{Cache miss}
		\If{$\mathsf{Q.size}=C$}
            \State $\mathsf{b' \leftarrow Q.pop()}$, \ $\mathsf{Cache.evict}(b')$ \label{ln:evict}
      \EndIf
		\State  $\mathsf{Cache.load}(b)$, \ $\mathsf{Q.push}(b, P[b][c[b]])$ \label{ln:main2}
      \EndIf
      \EndFor
	\end{algorithmic}}
\end{algorithm}

Algorithm~\ref{alg:belady} shows how \name implements Belady's algorithm. We use a max-heap $\mathsf{Q}$ to manage the cache-resident buckets, and each element of the heap is a key-value pair, where the key is the bucket id and the value is the index of the bucket's \emph{subsequent} access in the bucket access sequence. 
$\mathsf{Q}$ can hold $C$ elements, with $C$ being the capacity of the cache. \name also uses a two-dimensional array $\mathsf{P}$ to record when each bucket will be accessed, and each row, $\mathsf{P}[b]$, records the indices in the bucket access sequence for bucket $b$. A counter array $\mathsf{c}$ records how many times each bucket has been accessed until now.  Algorithm~\ref{alg:belady} fills in the index array $\mathsf{P}$ by enumerating the bucket access sequence $S$ (lines~\ref{ln:initP}--\ref{ln:initP2}). Then, the algorithm handles access one bucket at a time (lines~\ref{ln:main1}--\ref{ln:main2}). If the required bucket $b$ is in the cache (i.e., $\mathsf{Cache.find}(b) = \mathsf{true}$), we simply update the index of the bucket in the max-heap $\mathsf{Q}$ (line~\ref{ln:cacheHit}). If the bucket is not in the cache and the cache has free slots, we load the bucket and add its entry to $\mathsf{Q}$ (line~\ref{ln:main2}). If the cache is full, we choose the bucket $b'$ with the largest next access index and remove it from both $\mathsf{Q}$ and the cache (line~\ref{ln:evict}). Algorithm~\ref{alg:belady} scans $S$ over two passes, and the worst-case complexity is $O(|E|\log C)$, since $|S|=2|E|$ and the size of the heap $\mathsf{Q}$ is $C$. 


\subsection{Task Ordering via Graph Reordering}\label{sec:method:reorder}
\begin{figure*}[!t]
    \centering
    \includegraphics[width=\textwidth]{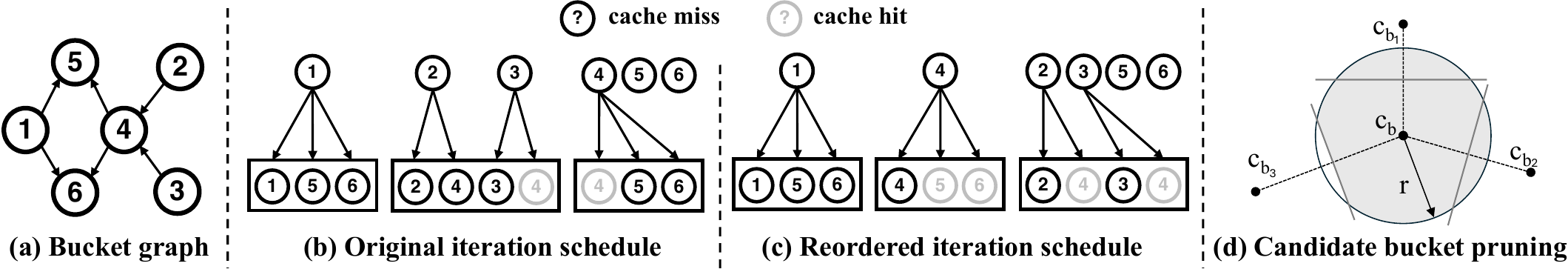}
    \vspace{-5mm}
    \caption{Given the bucket graph (a), the original task ordering (b) results in 8 cache misses in a cache that holds 3 nodes (rectangle box).  The reordered schedule (c) reduces the cache misses to 6. In the illustration of candidate bucket pruning (d), bucket $b_3$ is pruned, and as a result, the neighbors in the white arc are missed.
    }
    \label{fig:reorder}
\end{figure*}


Now we revisit the issue of determining the processing order of the edges in $G(V,E)$. 
Given a bucket node $v\in V$, let $\mathcal{N}(v)$ be the set of $v$'s out-neighbors, i.e., $\mathcal{N}(v)=\{u : (v,u)\in E\}$.  There are $|\mathcal{N}(v)|$ edges between $v$ and $\mathcal{N}(v)$, and they all share bucket $v$.  The first intuition applied in our heuristic is to process all of $v$'s outgoing edges in succession.  This ensures that at least one bucket for each task will not incur a cache miss.  When the cache size is $C\ge 2$, this reduces the worst-case cache miss count from $2|E|$ to $|V|+|E|$. As $|E|$ is usually much larger than $|V|$, the reduction is close to 50\%.

Now, task ordering boils down to determining the processing order of the nodes in $G$. While following the order of the node IDs presents a straightforward solution, it may still incur a high number of cache misses. For instance, \autoref{fig:reorder}(b) shows that following the ID order needs to load 8 nodes to the cache. In comparison, \autoref{fig:reorder}(c) shows that by processing node 4 before nodes 2 and 3, the number of loaded nodes can be reduced to 5. This is because nodes 1 and 4 share common neighbors $\{5,6\}$, and processing node 4 immediately after node 1 can reuse $\{5,6\}$ in the cache.

\looseness-1
The example in \autoref{fig:reorder} suggests that we should process nodes $u$ and $v$ subsequently if the overlap between their neighbors, $\mathcal{N}(u)$ and $\mathcal{N}(v)$, is high. This allows for the reuse of nodes already loaded to the cache. More generally, denote the average out-degree of the nodes in $G$ as $d_{avg}$, a cache with capacity $C$ can hold the neighbors of $C/d_{avg}$ nodes. We should decide an ordering for the nodes in $G$ such that the nodes in a sliding window of size $w=C/d_{avg}$ can share the most neighbors. Thus, we aim to maximize the following score.
\begin{equation}
    {\small
	\begin{aligned}
	F(P)=\!\!\!\!\!\sum_{0<P(v)-P(u)\leq w}\!\!\!\!\!\!\!|\mathcal{N}(u)\cap \mathcal{N}(v)|
	=\sum_{i=1}^{n}\sum_{j=\max\{1,i-w\}}^{i-1}\!\!\!\!\!\!\!|\mathcal{N}(u)\cap \mathcal{N}(v)|,
	\end{aligned}}
\end{equation}
where $n$ is the number of nodes and $P:[1,n]\rightarrow[1,n]$ is an one-to-one function that maps a node from its new id to original id. We will process the nodes following their new ids.     


\begin{algorithm}[!t]
	\caption{Task Ordering with Graph Reordering} 
	\label{alg:reorder}
    {\small
	\begin{algorithmic}[1]
		\Require Bucket graph $G(V,E)$, sliding window size $w=\frac{C}{d_{avg}}$
		\Ensure An ordering $P$ for the nodes in $G$
		\State select a node $v$ as the start node, $P[1] \leftarrow v$ \label{ln:first-node}
		\State $V_R \leftarrow V(G) \backslash \{v\}$, $i \leftarrow 2$ \label{ln:greedy1}
		\While{$i \leq n$} \Comment{Greedily insert remaining nodes to P}
		\State $v_{max} \leftarrow \max_{v\in V_R} \sum_{j=\max\{1,i-w\}}^{i-1}\big| \mathcal{N}(P[j])\cap \mathcal{N}(v)\big|$ 
		\State  $P[i] \leftarrow v_{max}, i \leftarrow i + 1$
		\State  $V_R \leftarrow V_R \backslash \{v_{max}\}$ \label{ln:greedy2}
      \EndWhile
		\State \Return  $P$
	\end{algorithmic}}
\end{algorithm}

The formulation above resembles Gorder~\cite{gorder}, a classical graph reordering algorithm that aims to improve the cache locality of in-memory graph processing. We use the greedy heuristic in Algorithm~\ref{alg:reorder} to conduct the graph reordering. In particular, to initialize, we select the node with the largest number of out-neighbors as the first node (line~\ref{ln:first-node}). Then, in each step, we select the nodes from the remaining node set $V_R$ that has the maximum overlap in its neighbors with the previous $w$ nodes in the sliding window (lines~\ref{ln:greedy1}--\ref{ln:greedy2}).
Calculating the overlap and selecting the maximum for every iteration has a prohibitive time complexity: $O(w\cdot d_{max}\cdot n^2)$ where $d_{max}$ is the maximum out-degree of $G$.
To resolve this, our implementation  adopts a priority queue to keep track of the following score $k_v$ for each node $v\in V$ 
$$k_v=\sum_{j=\max\{1,i-w\}}^{i-1}\big| \mathcal{N}(P[j])\cap \mathcal{N}(v)\big|$$

\noindent which is the number of overlapping neighbors between $v$ and nodes in the current sliding window. We incrementally update only the $k_v$ of nodes $v$ that are influenced by $v_o$ and $v_i$---the nodes going out of and into the window, respectively, while the window is sliding. The algorithm's time complexity is $O(\sum_{u\in V}(d^+(u)^2))$, where $d^+(u)$ is the out-degree of  $u$. Our evaluation shows that our optimized implementation incurs little overhead, as $d^+(u)$ is usually not large. 

\stitle{Putting the two components together} 
Task orchestration first runs graph reordering to determine the processing order of the nodes, which induces a processing order of the edges. During task execution, we follow the order to process the edges and use Algorithm~\ref{alg:belady} for cache management.

\section{Bucketing Optimizations}\label{sec:opt}

The core of \name relies on bucket-wise processing of SSJs.  In this section, we discuss how the partitioning of the vector dataset into buckets can be done effectively and efficiently, and detail our pruning strategy when building the bucket graph.


\subsection{Efficient Vector Bucketization}\label{sec:opt:bucketing}


\rfour{
During vector bucketization, \name groups the vectors of the dataset $\mathcal{X}$ into buckets that contain similar vectors. Standard clustering approaches cannot be straightforwardly adapted for our setting.  For example, K-means would require a full disk dataset scan per iteration for assigning points to clusters and updating the centroids.  This overhead across multiple iterations is impractical in our memory-constrained setting.
Here, we propose an efficient 3-step vector bucketization process that respects strict memory constraints (e.g., 5\% of the dataset size). 
}


\looseness-1
First, we select a set $\mathcal{X}'$ of randomly-sampled vectors as the bucket centers. We first generate the ids of the vectors to sample and then stream the entire dataset over memory to collect the sampled vectors. If the vectors in $\mathcal{X}$ are already permuted, we may simply use the first $|\mathcal{X}'|$ vectors to save disk traffic. The number of centers and, hence, buckets is configured to be large (e.g., around 1\textperthousand\ of the dataset size); this leads to small buckets and fine-grained space partitioning, which reduces the number of vectors each bucket needs to check for $\epsilon$-neighbors and, hence, disk traffic. A random sample is sufficiently representative of the dataset because the sample size is large (e.g., a 1\textperthousand\ sample of a billion-scale dataset still contains 1M vectors).
\rfour{This step safely meets the memory constraint since it only requires streaming the dataset and preserving 1\textperthousand\ of it in memory.}

Then, we read a block of vectors from the disk each time to avoid read amplification, and for each vector $x$ in the block, we search for its nearest center $c$ in $\mathcal{X}'$ for bucket assignment. A brute-force scan over $\mathcal{X}'$ for this purpose is expensive as $\mathcal{X}'$ is also a large vector dataset. As such, we construct an HNSW~\cite{hnsw} on $\mathcal{X}'$ in memory, which is one state-of-the-art index for vector similarity search (VSS), and use $x$ as a query to search the HNSW to identify its nearest center $c$. In particular, HNSW is a proximity graph index, where vectors are the graph nodes and edges connect similar vectors. A VSS query is processed by a traversal on the proximity graph that moves toward the neighbors with small distances to the query. HNSW can be built with complexity $O(|\mathcal{X}'|\mathsf{log}|\mathcal{X}'|)$, and process a VSS query with $O(\mathsf{log}|\mathcal{X}'|)$,  which is much faster than linear scan. \rfour{The memory footprint of the HNSW index---including the vectors in $\mathcal{X}'$ and the adjacency lists---is only about 2\textperthousand\ of the dataset size. Therefore, it comfortably fits within reasonable memory constraints.}

We do not write $x$ immediately to $b_c$ (i.e., the bucket corresponding to center $c$) as each vector is usually smaller than the 4KB disk page size and this causes write amplification. Instead, we keep an in-memory buffer for each bucket, append its vectors to the buffer, and flush the buffer to disk when it is full.

\rfour{
Our vector bucketing procedure is time-efficient because it only involves sequential disk accesses and requires minimal number of disk dataset scans---one for centroids sampling, one for bucket assignment and one for writing buckets back to disk. It is also memory-efficient since the minimum memory requirement is only 2\textperthousand\ of the dataset size, which can fit within any reasonable memory constraint.  
}
When it finishes, the vectors of $\mathcal{X}$ are stored as buckets on disk. 
\rthree{We note that the disk organization and the HNSW index on bucket centers are independent of the distance threshold $\epsilon$ and recall target $\lambda$. They can be built once and reused across queries with varying distance thresholds and recall requirements.}
The HNSW index for the centers is also used to build the dependency graph $G(V,E)$ for the buckets. In particular, for each bucket $b$ with center $c$, we use $c$ as a query to search the HNSW and identify a set $\mathcal{B}_c=\{b_1,b_2,\cdots,b_L\}$ of  nearest centers (i.e., buckets). Then, we use a probabilistic pruning rule (discussed next) to determine the buckets that should be checked by $b$ for $\epsilon$-neighbors. 

\subsection{Probabilistic Pruning for Candidate Buckets}\label{sec:opt:pruning}


Using the triangle inequality in Eq.~\eqref{equ:trangle} to decide the candidate buckets for each bucket $b$ will produce many candidate buckets. This will result in high disk access and computation costs as each bucket reads and checks many buckets. Inspired by~\cite{pruning}, we use a probabilistic candidate bucket pruning method to reduce the number of candidate buckets while meeting the recall target $\lambda$.   

\autoref{fig:reorder}(d) illustrates the idea with a toy example. We are considering bucket $b$ and have identified all the candidate buckets for verification, i.e.,  $\{b_1, b_2, b_3\}$, whose centers are plotted as black points. We also plot a ball with center $c_{b}$ and radius $r$, which represents the $\epsilon$-neighborhood of bucket $b$. For each candidate bucket (e.g., $b_1$), we plot its boundary region with the $\epsilon$-neighborhood of bucket $b$, which is a mid-vertical hyperplane as we assign each vector to its nearest center. We prune the candidate buckets in descending order of their distances to $c_{b}$. As shown in the example, if bucket $b_3$ is pruned, any $\epsilon$-neighbors in the white arc will be missed. We assume that the vectors are uniformly distributed in the hypersphere $B(c_{b}, r)$, and estimate the percentage of missed neighbors based on relative volume. We can keep pruning candidate buckets until the estimate of missed neighbors reaches $1-\lambda$.


\begin{algorithm}[!t]
	\caption{Probabilistic Pruning for Candidate Buckets} 
	\label{alg:prune}
    {\small
	\begin{algorithmic}[1]
		\Require Target recall $\lambda$, candidate bucket list $L$ for bucket $b$
		\Ensure Pruned candidate bucket list $L'$
		\State $L' \leftarrow L$, \ $\mathsf{sum} \leftarrow 0$
		\For{$b_i\in L$ in descending order of distance to $b$}
		\State $\mathsf{sum} \leftarrow \mathsf{sum} + \mathsf{arccos}(\mathsf{min}\{x_i,1\})$
		\If{$\mathsf{sum} \ge 1-\lambda$} \Comment{Stop when error exceeds $1-\lambda$}
		\State  break
      \EndIf
		\State $L' \leftarrow L' \backslash \{b_i\}$
      \EndFor
		\State \Return  $L'$
	\end{algorithmic} }
\end{algorithm}

More generally, assume that a bucket $b$ has $l$ candidate buckets, and we prune the $j$ furthest buckets from $b$. The percentage of missed neighbors can be expressed as  
\begin{align}\label{equ:prune final}
    \beta(j)&=\frac{V(\bigcup_{i=l-j}^{l}A(i))}{V(B(c_{b}, r))} \leq \frac{\sum_{i=l-j}^{l}V(A(i))}{V(B(c_{b}, r))}
\end{align}
where $V(\cdot)$ is the volume of a high-dimension shape, $A(i)$ is the arc between bucket $b$ and bucket $b_i$. It has been shown by~\cite{pruning} that $\beta(j)$ can be bounded by
{\small
\begin{align}
    \beta(j)&\leq \pi^{-\frac{1}{2}}\frac{\Gamma(\frac{d-1}{2})}{\Gamma(\frac{d}{2})}\sum_{i=l-j}^{l}\mathsf{arccos}(\min\{x_i,1\})
    =\mu\sum_{i=l-j}^{l}\mathsf{arccos}(\min\{x_i,1\}) \notag
\end{align}}
where $\mu=\pi^{-\frac{1}{2}}\frac{\Gamma(\frac{d-1}{2})}{\Gamma(\frac{d}{2})}$ is a constant related to the dimension $d$, and $\Gamma{\cdot}$ is the gamma function. 
$x_i=\frac{db_i}{r}$, where $db_i$ is the distance from point $c_b$ to the line  the boundary of $b,b_i$. When $x_i > 1$, $b$ and $b_i$ have no intersection, and thus does not contribute to the sum.

Algorithm~\ref{alg:prune} shows how to use Eq.~\eqref{equ:prune final} to prune the candidate buckets. Given a desired recall $\lambda$, for each bucket $b$, we sort its candidate buckets in decreasing order of their distances to $b$, and take the sum of $arccos(x)$ starting from the most distant bucket until the sum reaches the error bound $1-\lambda$. We stop and prune the buckets encountered so far.

%% file: sections/5-evaluation.tex
\begin{figure}
\begin{tabular}{lcccc}
\toprule
\textbf{Name}   && \textbf{Vectors}     & \textbf{Dimensions} & \textbf{Size (GB)}      \\ 
\midrule
Deep100M        && 100,000,000 & 96 & 36   \\
BigANN100M       &      & 100,000,000          & 128         & 48           \\
SPACEV1B          &   & 1,402,202,072          & 100          & 523           \\ 
\bottomrule
\end{tabular}
\vspace{-2mm}
\caption{Summary statistics of the experiment datasets}
\label{tab:datasets}
\end{figure}

\begin{figure}[!t]
	\centering
	\includegraphics[width=0.45\textwidth]{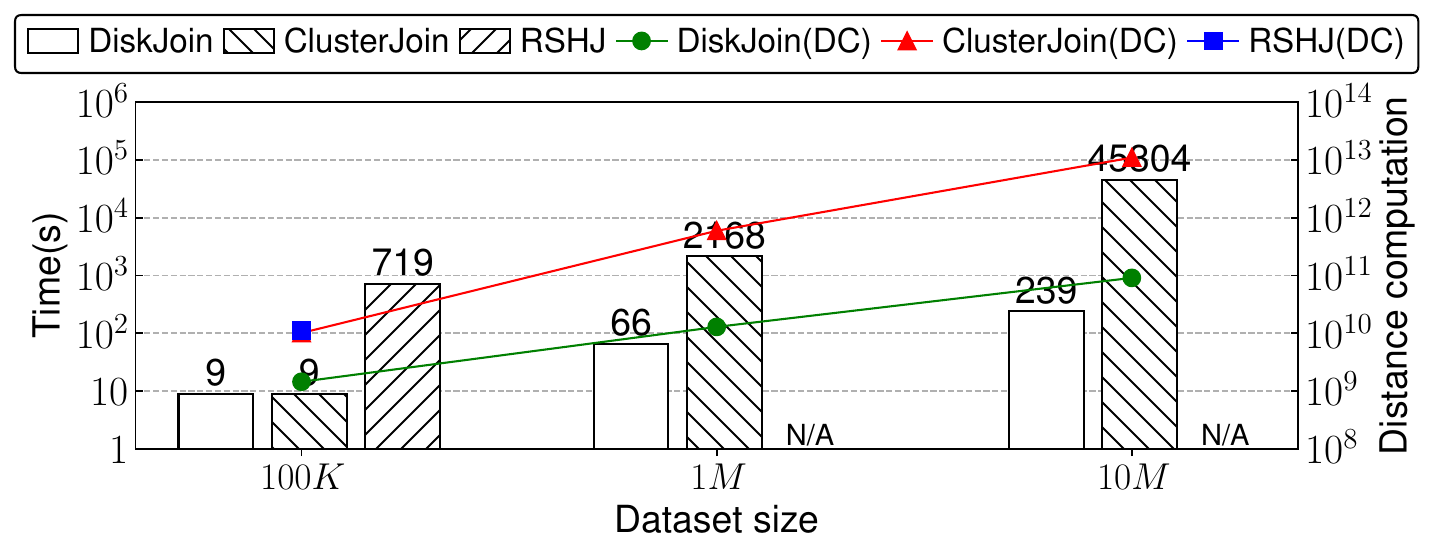}
    \vspace{-4mm}
	\caption{Despite being a disk-based method, \name is order of magnitudes faster than Clusterjoin and RSHJ---the latter failing to execute in larger data sizes. 
 This is because \name performs orders-of-magnitude fewer distance computations.
    }
	\label{exp:clusterjoin}
	\Description{}
\end{figure}

\begin{figure*}[!t]
	\centering
        \includegraphics[width=0.95\textwidth]{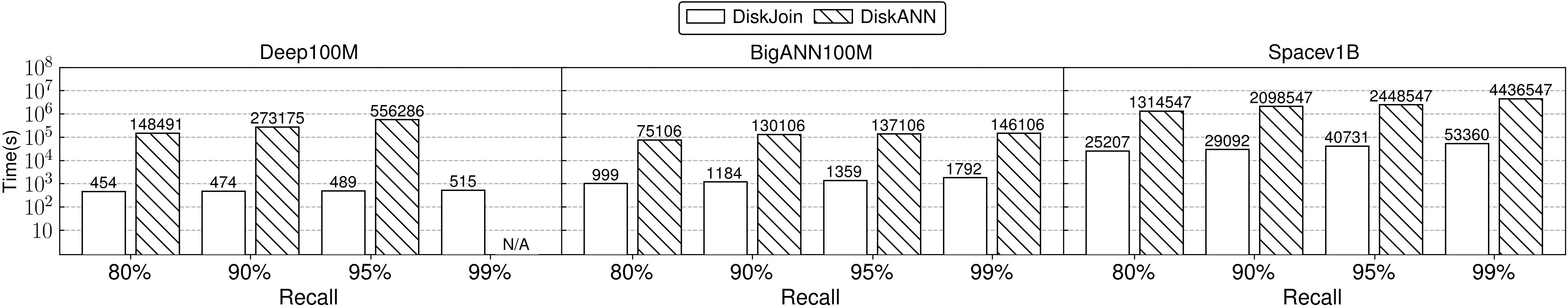}
        \vspace{-4mm}
	\caption{We set memory to 10\% of the dataset size and vary the target recall. \name is consistently 2-3 orders of magnitude faster than DiskANN.   DiskANN cannot
achieve 99\% recall for Deep100M due to the approximation errors in vector compression.}
	\label{exp:recall-time}
	\Description{}
    \vspace{-2mm}
\end{figure*}

\section{Experimental Evaluation}\label{sec:eval}

In this section, we present an extensive evaluation of \name, using three large vector datasets and three alternative baseline methods based on prior art. Our evaluation aims to answer the following questions:
\squishlist
\item \textit{How efficient is \name compared to related prior art, when the memory is small with respect to the dataset? }

\item \textit{How do the task configurations (i.e., target recall, distance threshold, and memory budget) affect the performance of \name? }

\item \textit{How effective are our designs (i.e., Belady's algorithm, task reordering, and pruning) in improving the efficiency of \name? }
\squishend

\begin{figure*}[!t]
	\centering
	\includegraphics[width=0.95\textwidth]{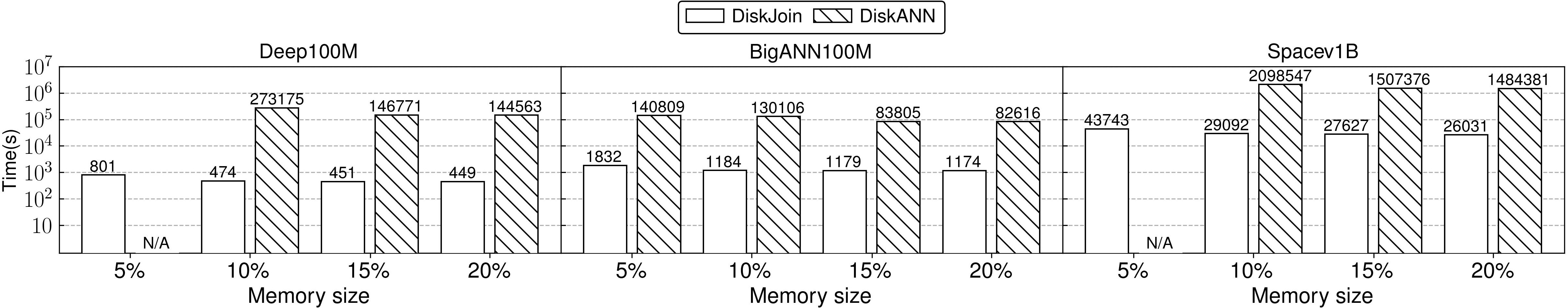}
    \vspace{-4mm}
	\caption{We set the target recall to 90\% and vary the memory budget (from 5\% to 20\% of the data size). \name is 2-3 orders of magnitude faster than DiskANN. In low memory settings, DiskAnn may fail to achieve the target recall due to low vector compression accuracy.}
	\label{exp:memory-time}
	\Description{}
    \vspace{-2mm}
\end{figure*}



\subsection{Experimental Settings}\label{sec:settings}

\stitle{Datasets} 
We use 3 datasets of varied characteristics. Deep100M (a subset of Deep1B~\cite{deep}) is a vector dataset of image embeddings produced by the last fully connected layer of the GoogLeNet model~\cite{googlenet}; it contains 100M vectors with 96 dimensions. BigANN100M contains 100M vectors that are scale-invariant feature transform (SIFT) descriptors of images, with 128 dimensions.  SPACEV1B consists of more than 1.4~billion 
vectors of $100$ dimensions, where each vector is a document embedding produced by the Microsoft SpaceV Superior model. The dataset statistics are summarized in \autoref{tab:datasets}.


\stitle{Baselines} To the best of our knowledge, no existing systems target disk-based vector similarity join. 
Therefore, we use the following prior art for related problem settings as baselines.
\squishlist
    \item DiskANN~\cite{diskann} is a popular state-of-the-art index for disk-based vector similarity search. As discussed in Section~\ref{sec:intro}, we first build a DiskANN index and then use each vector in the dataset as a query to search the index for $\epsilon$-neighbors. DiskANN is designed to search the top-$k$ similar neighbors for a query, and we adjust the value of $k$ to control the recall for  vector similarity join. Other disk-based vector indexes (e.g., Starling~\cite{starling}) may outperform DiskANN for \emph{some} datasets. However, the speedups they report compared to DiskANN are modest.  Given that  \name demonstrates much more significant speedups over DiskANN (50x) for similarity join, such disk-based vector index baseline alternatives do not alter our conclusions.
    \item ClusterJoin~\cite{clusterjoin} is a distributed similarity join solution. We implement a single-node in-memory version for fair comparison.
    \item RSHJ~\cite{lshsj} is an in-memory vector similarity join solution using locality-sensitive hashing (LSH).
\squishend

\stitle{Platforms} We conduct the experiments on two AWS servers, one server is equipped with a 48-core Intel Xeon Platinum 8375C CPU @ 2.90GHz, 96GB RAM, and 2 $\times$ 1.3TB NVMe SSDs, while the other server has a 72-core Intel Xeon CPU E5-2686 v4 @ 2.30GHz, 512GB RAM, and 8 $\times$ 1.8TB NVMe SSDs. We use the first server for all experiments on the Deep100M and BigANN100M datasets, and the second server for all experiments on the SPACEV1B dataset.

\stitle{\rone{Implementation}} \rone{We implement \name as a separate engine, independent from any RDBMSs. Future integration into RDBMS is certainly possible, and our algorithms would easily translate in this context.} \rthree{To prevent interference of OS page caching, we enabled the O\_DIRECT flag in all file manipulation operations.}

\stitle{Evaluation protocol} We are interested in the efficiency of vector similarity join and thus use \rthree{the end-to-end} execution time\rthree{---which includes any necessary index construction time (e.g., for \name and DiskANN)---}as the main performance metric. We restrict the memory usage to be a small portion (e.g., 5\%-20\%) of the vector dataset size \rone{for two purposes: (1)~it ensures that the compared methods use the same amount of memory, enabling a fair comparison and (2)~it allows us to emulate scenarios where vector datasets are so large that they exceed available memory, which is common in practice.}
By default, we use a memory that is 10\% of the dataset size, configure the target recall as $\lambda=0.9$, and set the distance threshold $\epsilon$ for similar vector pairs such that each vector has 100 similar vectors on average.

\subsection{Comparison with ClusterJoin and RSHJ}
\looseness-1
We first evaluate the performance of \name against the other similarity join baselines.
\autoref{exp:clusterjoin} reports the execution time and number of distance computations (DC) for \name, ClusterJoin, and RSHJ on BigANN dataset subsets of varying sizes: 100K, 1M and 10M. As ClusterJoin is an exact algorithm, we configure \name to achieve a recall of $99.5\%$ for a fair comparison. The recall of RSHJ, despite our best efforts to tune its parameters, reaches only 97.6\% and cannot be improved further. The memory budget of \name is constrained to $10\%$ of the dataset size. RSHJ fails to run on the 1M and 10M datasets due to its prohibitive $O(n^2)$ memory consumption. Despite being a disk-based approach, \name significantly outperforms the in-memory ClusterJoin and RSHJ. RSHJ performs particularly poorly due to a sub-procedure with $O(n^3)$ time complexity. Beyond that, the performance gap is primarily attributed to the substantial difference in the number of candidate pairs for distance computation in the verification phase. Notably, the number of distance computations in \name grows linearly with dataset size, whereas ClusterJoin exhibits near-quadratic growth. Considering the high overhead of network communication and disk I/O in a MapReduce environment, the distributed execution of CluserJoin is unlikely to outperform \name, while incurring higher resource costs.

\smallskip
\noindent
\fbox{
\parbox{0.96\columnwidth}{
\emph{Key takeaway:} 
   \name is significantly faster than in-memory state-of-the-art methods for similarity join, because it reduces distance computation by orders of magnitude.
}}

\subsection{
Performance under Varied Configurations}

In this section, we evaluate the performance of \name with respect to all its parameters.  We compare with DiskANN, which is the only disk-based method in our baselines.

\stitle{Target recall} \autoref{exp:recall-time} reports the execution time of \name and DiskANN when changing the target recall. Note that the y-axis is in log scale. Since DiskANN usually can not finish execution within one day, we estimate its execution time by sampling a subset of the vectors as queries. In particular, we randomly sample 1\textperthousand\ of the vectors, and this yields 100K-1M vectors for the datasets, which make the sample sets sufficiently large for accurate time estimation. The execution time of DiskANN is omitted at $99\%$ recall on Deep100M because DiskANN uses Product Quantization (PQ)~\cite{ivf} to compress vectors and stores the compressed dataset in memory. Under our default $10\%$ memory size constraint, DiskANN can not reach a recall of $99\%$ for Deep100M due to the approximation errors introduced by vector compression. 

This experiment shows that \name significantly outperforms DiskANN for all datasets and target recalls. In particular, the speedup of \name over DiskANN is $52\times$ at the minimum and $1137\times$ at the maximum (i.e., on Deep100M at a recall of 95\%). As we will show later with detailed profiling, this is because DiskANN suffers from long disk I/O time and computation time while \name's optimizations for disk I/O and candidate pruning effectively reduce both IO and computation costs. Both DiskANN and \name run longer to achieve higher target recalls. This is because \name needs to check more bucket pairs, and DiskANN needs to search more neighbors for each query. However, the execution time of DiskANN increases faster than that of \name. For example, to improve the recall from 80\% to 99\% for Spacev1B, execution time increases $4\times$ for DiskANN but only about $2\times$ for  \name.

\begin{figure}[t]
	\centering
		\includegraphics[width=0.8\columnwidth]{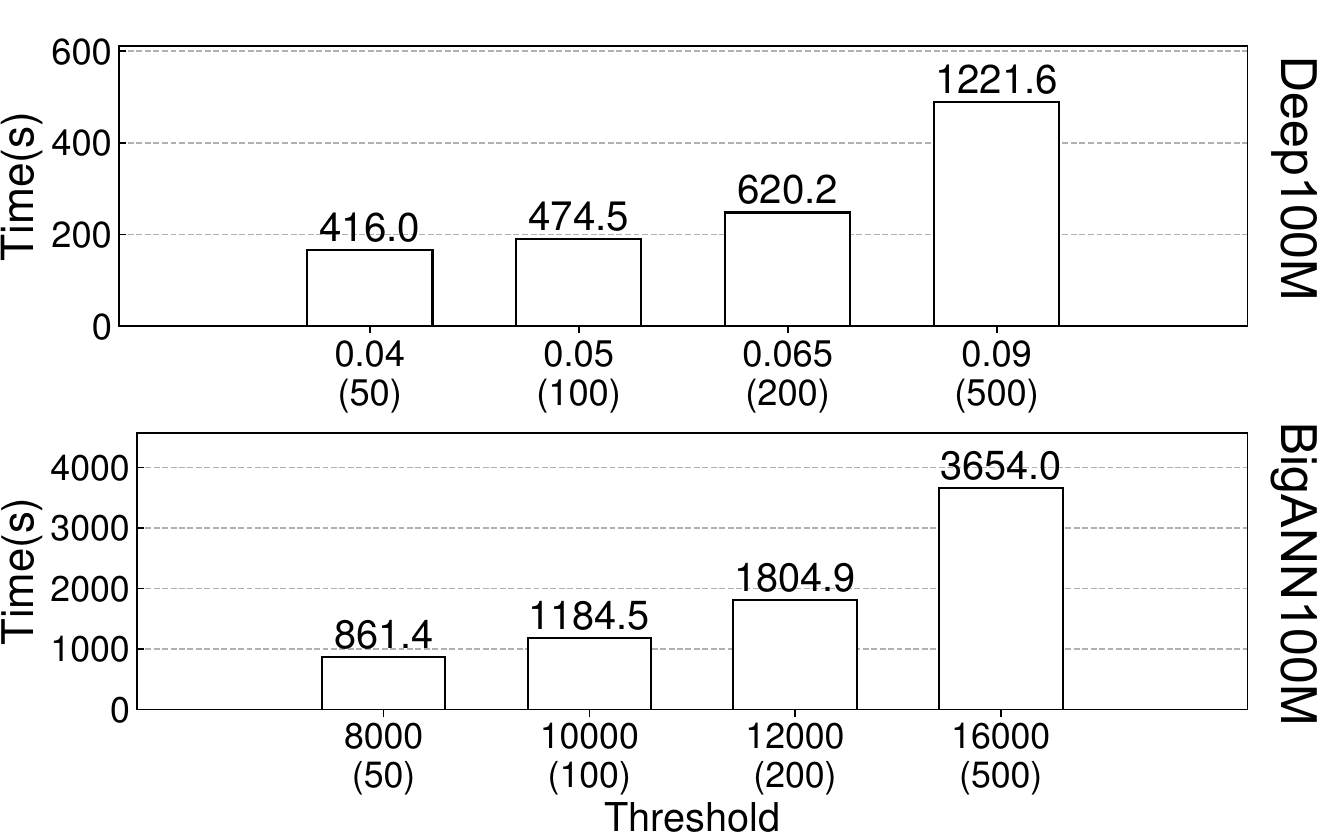}
	\vspace{-3mm}
	\caption{
    Execution time of \name increases as we increase the distance threshold $\epsilon$, but growth remains sublinear for settings up to 500 similar neighbors, on average.
    }
	\label{exp:epsilon}
\end{figure}

\stitle{Memory constraint} \autoref{exp:memory-time} reports the execution time of \name and DiskANN under varying memory constraints (from 5\% to 20\% of the data size), for fixed target recall (90\%). 
For \name, the memory constraint determines the number of buckets that can be cached in memory. For DiskANN, two factors affect its memory consumption: the compression ratio of the PQ for approximate vectors and the portion of the proximity graph index that is cached in memory. We first select the PQ compression ratio that is  closest to the memory constraint, and cache the proximity graph in the remaining memory. This is because DiskANN uses the approximate vectors for proximity graph traversal, and thus the compressed vectors need to be accurate for good recall. 
DiskANN fails to reach $90\%$ recall on Deep100M and Spacev1B when the memory constraint is $5\%$ of the dataset size due to vector compression errors.

\looseness-1
We observe that \name is consistently 2-3 orders-of-magnitude faster than DiskANN. We observe diminishing performance gains when enlarging the memory constraint beyond $10\%$, i.e., the speedup of increasing memory size from $10\%$ to $20\%$ is much smaller than increasing the memory size from $5\%$ to $10\%$. As we will show in later experiments, this is because at a memory size of $10\%$ dataset size, \name's execution time is dominated by computation rather than IO. Thus, further increasing the memory to reduce IO has a limited effect on overall execution time. This indicates a significant benefit of \name: it only needs a small amount of memory (with respect to the dataset) to avoid making IO the bottleneck. For DiskANN, the performance improvement is primarily attributed to the enhanced accuracy of the compressed vectors. For example, on Deep100M, increasing the memory from $10\%$ to $15\%$ of the dataset size results in a speedup of 1.86$\times$ due to more accurate compressed vectors. However, increasing the memory from $15\%$ to $20\%$ has nearly no reduction in execution time, as the vectors are already accurate, and the additional memory is used to cache the proximity graph index. 


\stitle{Distance threshold} 
In the next experiment, we vary the distance threshold $\epsilon$, so that each vector has 50, 100, 200 and 500 similar neighbors, on average.
\autoref{exp:epsilon} shows that the execution time of \name increases more quickly for BigANN100M than Deep100M. This is in line with the results in \autoref{exp:recall-time}, i.e., when increasing the target recall, \name's execution time grows slower for Deep100M than BigANN100M, suggesting that finding neighbors is more difficult in BigANN100M. Overall, growth remains sublinear for settings up to 500 similar neighbors, on average.

\begin{figure}[!t]
	\centering
	\includegraphics[width=0.45\textwidth]{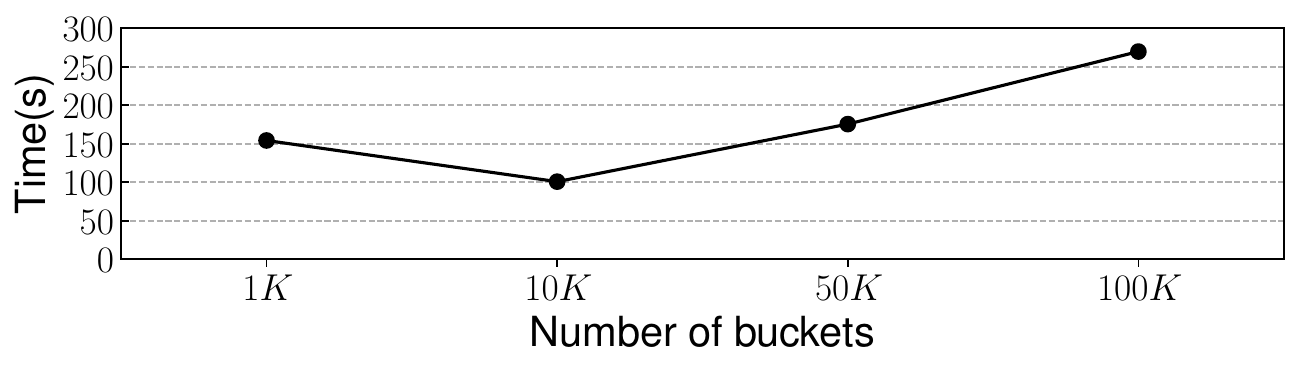}
    \vspace{-4mm}
	\caption{The number of buckets affects the performance of \name over a subset of BigANN100M with 10M vectors. The best time is achieved at around 1\textperthousand\ of the data size.
    }
	\label{exp:bucket-number}
	\Description{}
\end{figure}

\begin{figure}[!t]
	\centering
	\includegraphics[width=0.45\textwidth]{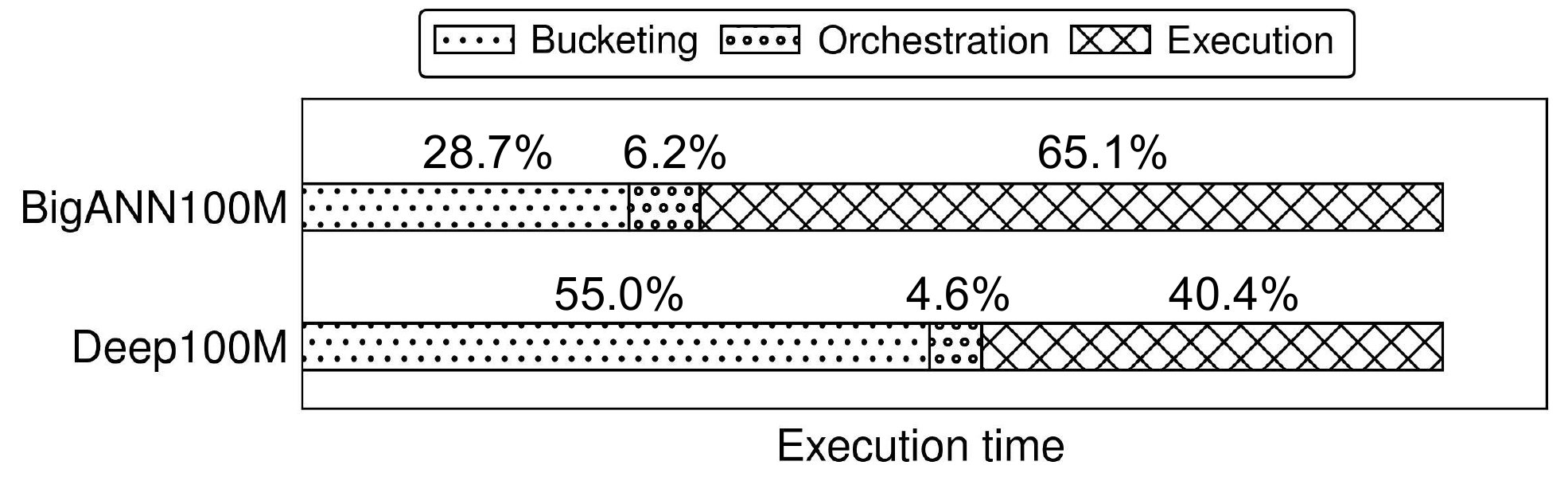}
    \vspace{-4mm}
	\caption{Execution time breakdown of \name into three parts: bucketing, orchestration and execution.}
	\label{exp:decomposition}
	\Description{}
\end{figure}

\stitle{Number of buckets} 
\autoref{exp:bucket-number} shows the execution time on a 10M size subset of BigANN100M, when varying the number of buckets used by \name as 0.1\textperthousand, 1\textperthousand, 5\textperthousand, and 1\% of the dataset size. \name achieves best performance when the number of buckets is around 1\textperthousand\ of the dataset size. With fewer buckets, the space partitioning is less fine-grained and thus produces more candidate pairs. Moreover, the number of buckets that can be cached decreases, leading to a higher cache miss rate and increased execution time. Conversely, when the number of buckets is large (e.g., 1\%), the average bucket size drops to around 100 vectors. Considering that bucket sizes are not evenly distributed, many buckets may occupy only a few pages or even less than a page, resulting in significant read amplification and degraded I/O efficiency.

\stitle{Execution time breakdown}
\autoref{exp:decomposition} reports the running time of the three main phases for \name: bucketing, orchestration, and execution. Bucketing maps the vectors to buckets and stores the buckets on disk; orchestration includes dependency graph construction, task reordering, and cache management; execution loads the buckets from disk and performs computations. The results show that the orchestration part, which encompasses our main optimizations, has a small overhead (around 5\% of the total execution time). For Deep100M, the time spent on bucketing exceeds that of execution. This is because most vector pairs are filtered out, leaving a small number of candidate pairs for verification.

\rfour{
\stitle{Randomness sensitivity} The execution time and recall depend on the quality of vector bucketization, which involves random center selection. To evaluate the robustness of this randomness, we repeat the experiment on the 10M subset of BigANN100M fifty times. The results show that performance is robust: the mean recall is 0.903 with a standard deviation of 0.005, and the mean execution time is 276.02 seconds with a standard deviation of 12.62 seconds. This low variance indicates that the performance is largely insensitive to the specific random centers chosen.
}

\rfour{
\stitle{Stress-testing \name} \name{} assumes ideal sequential disk layout. In practice, however, files may split into non-contiguous pieces (extents) as the file system ages, a phenomenon known as file system fragmentation. In this experiment, we stress-test \name under such degraded disk conditions. We create three 100GB file systems, fill them with small files of varying sizes (16KB, 128KB and 1024KB), and then delete a random subset of files to simulate fragmentation. Figure~\ref{exp:stress-test} shows the execution time of \name and the number of extents in the disk index file on the 10M size subset of BigANN100M across different fragmentation level. As fragmentation increases, the disk index is broken into more extents, increasing execution time. However, the performance impact remains minimal at moderate fragmentation level (1024KB and 128KB) and only becomes noticeable---still under a 10\% slowdown---under severe fragmentation. This is due to the fact that SSDs do not physically seek, so there is no difference between non-sequential access and sequential access. The performance of \name only degrades when the file system is so fragmented that many extents of the disk index have size less than 4KB, causing read amplification.
}

\begin{figure}[!t]
	\centering
	\includegraphics[width=0.45\textwidth]{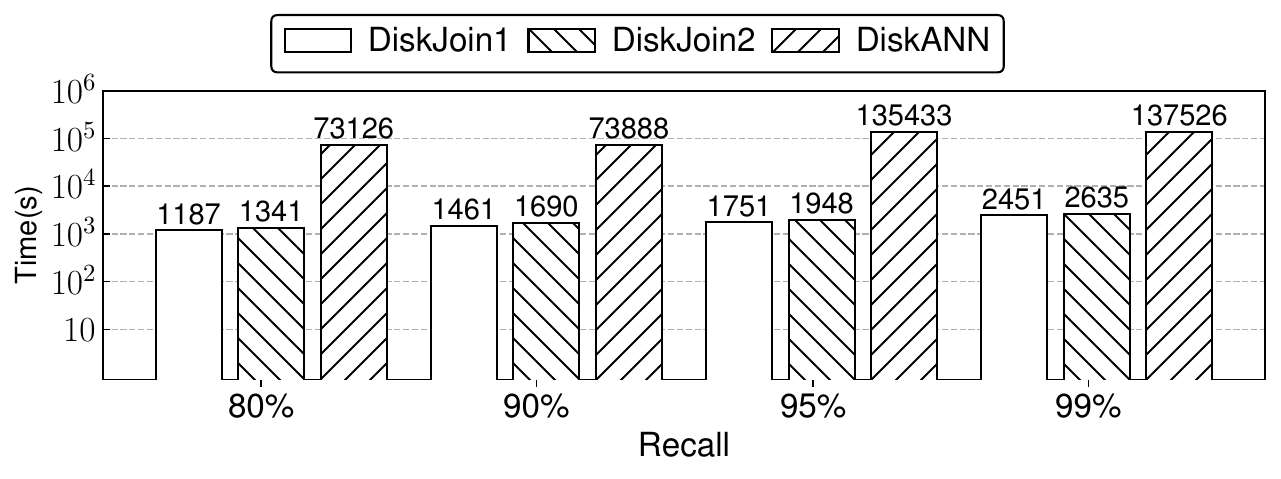}
    \vspace{-3mm}
	\caption{Execution time of cross-join on two Deep datasets of size 50M and 100M. DiskJoin1 denotes reordering the larger dataset and caching the smaller dataset, while DiskJoin2 means the opposite.}
	\label{exp:cross-join}
	\Description{}
\end{figure}

\begin{figure}[!t]
{\small\rfour{
\begin{tabular}{ccc}
\toprule
\textbf{Fragmentation level} & \textbf{No. of extents} & \textbf{Execution time (s)} \\ \midrule
None                & 536            & 108.1              \\ 
1024KB              & 1579           & 110.7              \\
128KB               & 6530           & 111.4              \\
16KB                & 265565         & 123.6              \\
\bottomrule
\end{tabular}}
}
\vspace{-2mm}
\caption{\rfour{Performance of \name is robust under moderate fragmentation conditions, with some degradation only when the number of extents becomes extremely large (at 16KB).}}
\label{exp:stress-test}
\end{figure}

\stitle{Cross-join}
We evaluate \name's performance in cross-joins using two disjoint subsets of the Deep1B dataset~\cite{deep}, of sizes 50M and 100M, respectively. The distance threshold and memory constraint are set to their default values. As discussed in Section~\ref{sec:overview}, \name may conduct cross-join in two ways: (1)~reordering the larger dataset and caching the smaller dataset or (2)~the reverse. These are denoted as \name{}1 and \name{}2 in \autoref{exp:cross-join}, respectively. Both \name{}1 and \name{}2 outperform DiskANN consistently, across different target recalls, and \name{}1 runs slightly faster than \name{}2. This is expected, as \name{}1 has shorter disk read time compared with \name2 (Section~\ref{sec:overview}).

\smallskip
\noindent
\fbox{
\parbox{0.96\columnwidth}{
\emph{Key takeaway:} 
   The execution time  of \name is consistently 2-3 orders of magnitude lower than DiskANN, and remains robust across varied configuration settings.
}}



\begin{figure}[!t]
	\centering
		\includegraphics[width=\columnwidth]{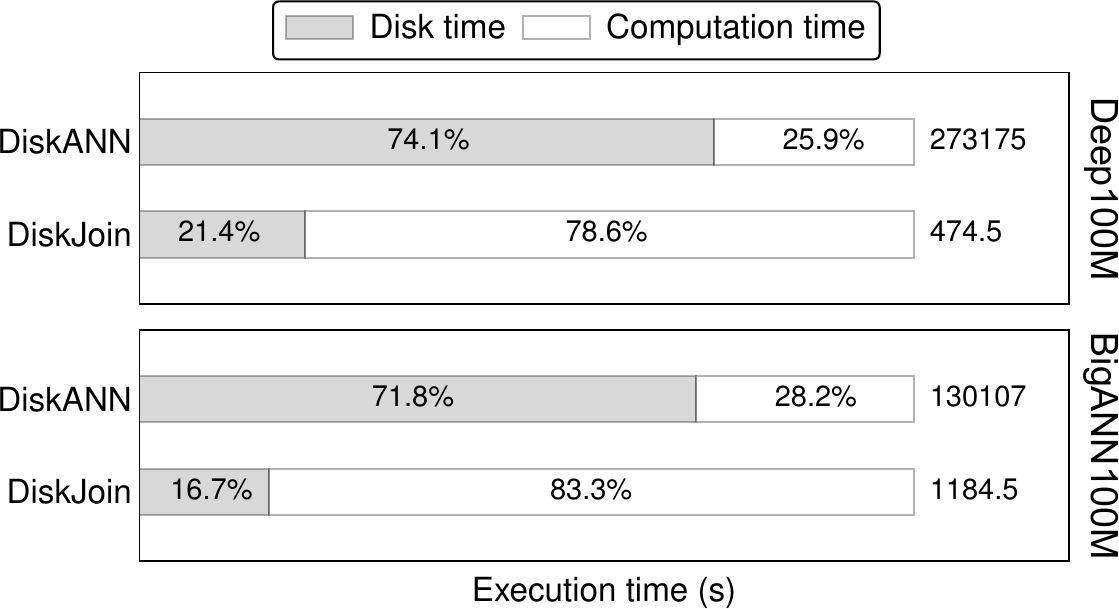}
	\vspace{-5mm}
	\caption{Breaking down the execution time for \name and DiskANN. Note that the lengths of the bar do not correspond to total time. 
    }
	\label{exp:breakdown}
	\Description{}
\end{figure}

\subsection{Performance Analysis}

In this section, we profile the execution statistics of DiskANN and \name to understand their performance. For all experiments, we use the default configurations (Section~\ref{sec:settings}).

\autoref{exp:breakdown} breaks down the execution time of DiskANN and \name into IO time and computation time. Note that the lengths of the bars do not indicate the execution time of the solutions, instead, we report the execution time next to the bars. The results show that disk IO takes up about $70\%$ of DiskANN's execution time on both datasets, while \name only spends $21.4\%$ and $16.7\%$ of the execution time on disk IO respectively.  
\autoref{tab:disk traffic} further explains this result, reporting the useful disk traffic against the total disk traffic for each approach (and their ratio, which is the read amplification). 
 We see that DiskANN needs to read far more data from the disk than \name. In particular, the average disk read volume per vector is 1.4MB and 1.2MB for DiskANN while \name only reads 1.9KB and 2.8KB. The large difference between \name's and DiskANN's disk traffic (i.e., over 400x) is due to three reasons. (1)~DiskANN suffers from severe read amplification due to the random access pattern of the proximity graph index. As a result, only a small fraction of the data read from disk is actually used by DiskANN. In contrast, \name enjoys nearly no disk amplification thanks to its bucket-based block disk access. (2)~\name optimizes for disk I/O through its dynamic memory cache and task reordering. (3)~\name's pruning technique successfully prunes most of the candidate pairs, further reducing the disk I/O time and computation time, which we will show later.

Overall, these results indicate that, with our optimizations, disk IO is no longer the bottleneck for vector similarity joins. Instead, the limiting factor becomes computation and the execution time of \name can be further accelerated by using GPUs for computation.   

\begin{figure}[!t]
\centering

{\small
\begin{tabular}{@{}cccccc}
\toprule
\multicolumn{2}{c}{\textbf{}} && Total (GB) & Useful (GB) & Amp. \\
\midrule
\multirow{2}{*}{\textbf{Deep}}    & DiskJoin  && 181     & 180.3   & 1.0039      \\
                                  & DiskANN   && 139,485 & 19,427  & 7.18      \\
\hline
\multirow{2}{*}{\textbf{BigANN}}  & DiskJoin  && 265     & 264.3   & 1.0026      \\
                                  & DiskANN   && 115,854 & 19,118  & 6.06      \\ 
\bottomrule
\end{tabular}}
\vspace{-2mm}
\caption{The disk traffic of \name and DiskANN. \textit{Amp.} is the ratio of total disk traffic over the useful disk traffic. 
}
\label{tab:disk traffic}
\vspace{-2mm}
\end{figure}

\smallskip
\noindent
\fbox{
\parbox{0.96\columnwidth}{
\emph{Key takeaway:} 
   Remarkably, \name achieves nearly 0 read amplification, eliminating disk IO as the bottleneck.
}}

\begin{figure}[!t]
	\centering
		\includegraphics[width=\columnwidth]{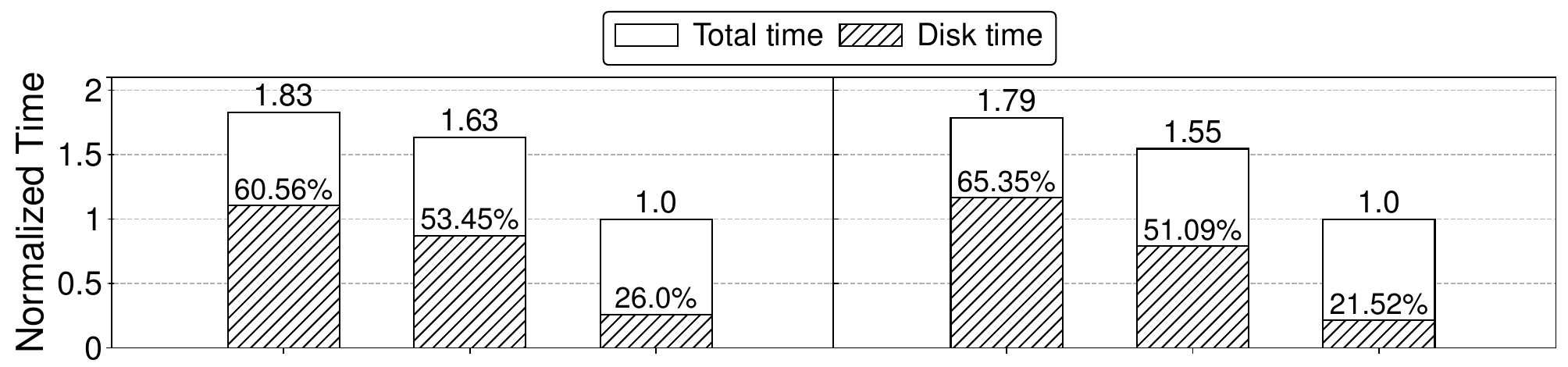}
		\includegraphics[width=\columnwidth]{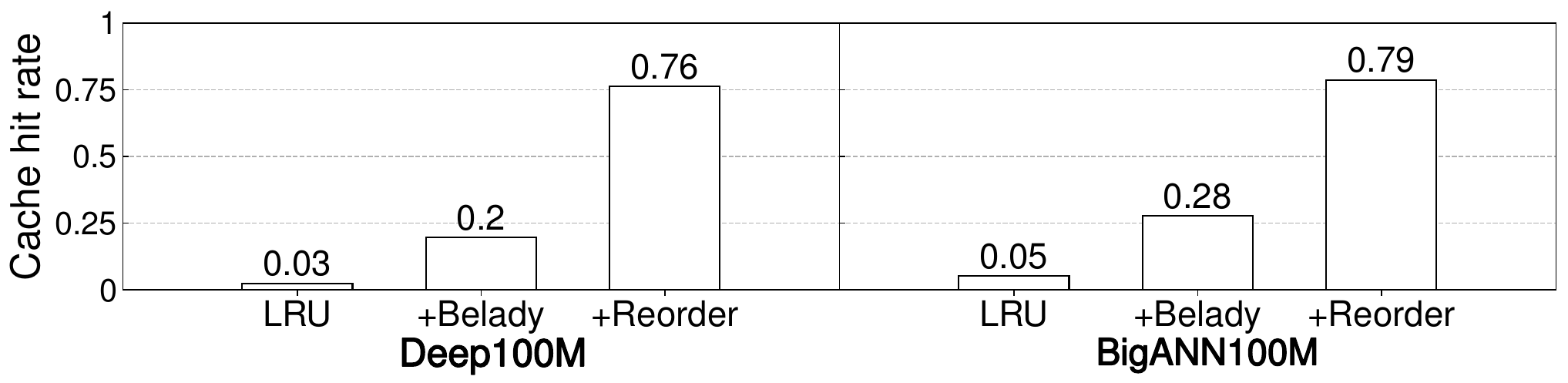}

	\vspace{-4mm}
	\caption{Normalized execution time and cache hit rate when enabling the task orchestration optimizations. 
    }
	\label{exp:cache}
	\Description{}
\end{figure}

\subsection{Effectiveness of \name Designs}

\rthree{In this section, we analyze the impact of our three optimizations: cache management using Belady’s algorithm, task reordering, and candidate pruning. Cache management and task reordering preserve the graph structure and thus do not reduce recall, while candidate pruning may affect recall, as it modifies the dependency graph. However, despite its heuristic nature, pruning performs very well empirically, demonstrating significant gains in efficiency with only minimal impact on recall (approximately a 0.1\% reduction in our experiments).  Below, we present an experimental analysis of how these optimizations influence efficiency.}

First, we consider our task orchestration optimizations. In \autoref{exp:cache}, the method annotated as $\mathsf{LRU}$ disables both cache optimizations, instead using the node id to determine task order and LRU as the cache replacement policy; $\mathsf{+Belady}$ uses Belady's algorithm instead of LRU; finally, $\mathsf{+Reorder}$ augments $\mathsf{+Belady}$ with task reordering.  For ease of comparison, we normalize the execution time results by the execution time of $\mathsf{+Reorder}$, and report the percentage of disk IO time within total execution time.


Both optimizations are effective at reducing the execution time of \name, as they improve the cache hit rate when accessing the buckets. When the two optimizations are disabled ($\mathsf{LRU}$), disk IO takes up more than half of the execution time, while disk IO is no longer the bottleneck when the optimizations are used ($\mathsf{+Reorder}$). Task reordering is particularly effective in improving the cache hit rate: When both optimizations are used, the cache hit rate is above 75\% for both datasets. This is remarkable because the memory size is only 10\% of the dataset size. The cache hit rate of $\mathsf{LRU}$ is below 10\% because similarity join does not have popular buckets, and a bucket that was recently accessed may not be accessed in the future, which constitutes an adversarial case for LRU.

\rfour{
Finally, we examine the effect of candidate bucket pruning, which directly influences the average degree of the bucket dependency graph---and consequently, the computational and disk I/O costs. \autoref{tab:pruning} reports the number of candidate vector pairs\footnote{\rfour{The number of candidate vector pairs is proportional to the number of candidate buckets---both reflect the computational and I/O costs; we report the more fine-grained statistic here.}} and execution time under varying distance thresholds. The results show that pruning leads to a substantial reduction in both the number of candidate vector pairs and execution time across all settings. For example, on the Deep100M dataset with a threshold of 0.05, pruning reduces the number of candidate pairs from $5.59 \times 10^{12}$ to $1.81 \times 10^{11}$, resulting in a drop in execution time from 5188.1 seconds to just 474.5 seconds. The trend holds for larger thresholds as well: at 0.09, pruning still reduces the execution time by nearly $9\times$ despite the increased candidate volume.
}


\begin{figure}[!t]
\centering
\resizebox{\columnwidth}{!}{
\rfour{
\begin{tabular}{c|c|cc|cc}
\toprule
\multirow{2}{*}{\textbf{Dataset}} & \multirow{2}{*}{\begin{tabular}[c]{@{}c@{}}\textbf{Threshold}\\(\# neighbors)\end{tabular}} & \multicolumn{2}{c|}{\textbf{No. of candidates}} & \multicolumn{2}{c}{\textbf{Execution time (s)}} \\
&& \multicolumn{1}{l}{w/o pruning} & \multicolumn{1}{l|}{w/ pruning} & \multicolumn{1}{l}{w/o pruning} & \multicolumn{1}{l}{w/ pruning} \\ \hline
\multirow{4}{*}{\textbf{Deep100M}}    
& \begin{tabular}[c]{@{}c@{}}0.04 (50)\end{tabular}   & $4.19\times 10^{12}$ & $1.12\times 10^{11}$ & \phantom{0}4,204 & \phantom{0}416 \\ \cline{2-6} 
& \begin{tabular}[c]{@{}c@{}}\phantom{0}0.05 (100)\end{tabular}  & $5.59\times 10^{12}$ & $1.81\times 10^{11}$ & \phantom{0}5,188 & \phantom{0}475 \\ \cline{2-6} 
& \begin{tabular}[c]{@{}c@{}}0.065 (200)\end{tabular} & $7.31\times 10^{12}$ & $3.93\times 10^{11}$ & \phantom{0}7,671 & \phantom{0}620 \\ \cline{2-6} 
& \begin{tabular}[c]{@{}c@{}}\phantom{0}0.09 (500)\end{tabular}  & $9.66\times 10^{12}$ & $1.14\times 10^{12}$ & 10,696 & 1,222 \\ \hline
\multirow{4}{*}{\textbf{BigANN100M}} 
& \begin{tabular}[c]{@{}c@{}}8000 (50)\end{tabular}   & $5.89\times 10^{12}$ & $4.60\times 10^{11}$ & \phantom{0}8,235 & \phantom{0}861 \\ \cline{2-6} 
& \begin{tabular}[c]{@{}c@{}}10000 (100)\end{tabular} & $7.76\times 10^{12}$ & $7.78\times 10^{11}$ & \phantom{0}9,930 & 1,185 \\ \cline{2-6} 
& \begin{tabular}[c]{@{}c@{}}12000 (200)\end{tabular} & $9.52\times 10^{12}$ & $1.26\times 10^{12}$ & 14,793 & 1,805 \\ \cline{2-6} 
& \begin{tabular}[c]{@{}c@{}}16000 (500)\end{tabular} & $1.28\times 10^{13}$ & $2.57\times 10^{12}$ & 20,813 & 3,654 \\ 
\bottomrule
\end{tabular}}
}
\vspace{-2mm}
\caption{
\rfour{Impact of candidate bucket pruning on candidate count and execution time. Pruning yields significant performance gains across all thresholds.}
}
\label{tab:pruning}
\end{figure}

\smallskip
\noindent
\fbox{
\parbox{0.96\columnwidth}{
\emph{Key takeaway:} 
   Our ablation study shows that all of \name's optimizations are useful in reducing disk access and runtime, \rthree{with minimal impact on recall.}
}}

%% file: sections/6-conclusion.tex
\section{Conclusion}

In this paper, we study the problem of similarity-based self-join over high-dimensional, large vector datasets.  We focus on disk-based solutions and observe that disk access dominates the execution time of disk-based vector similarity join due to read amplification and repetitive data access. We propose \name, which relies on smart bucket-wise processing of SSJs to reduce disk IO.
\name  carefully manages the in-memory data cache to improve cache hit rate, which is achieved by ordering the tasks for temporal locality and using a cache eviction policy that fully utilizes data access information. Through extensive experiments, we show that \name achieves consistently 2-3 orders-of-magnitude improvement in execution time and all but eliminates read amplification. Remarkably, \name removes disk access as a bottleneck in vector similarity self-join.